\documentclass[a4paper,11pt]{article}
\pdfoutput=1 % if your are submitting a pdflatex (i.e. if you have
\usepackage{jheppub} % for details on the use of the package, please
\usepackage{amsmath,amssymb,amsthm,amscd,graphicx}
\usepackage{psfrag}
\input epsf.sty

\usepackage{color}

\theoremstyle{definition}

\newcommand{\CC}{{\cal C}}
\newcommand{\cD}{{\cal D}}

\newcommand{\CF}{{\cal F}}
\newcommand{\CG}{{\cal G}}

\newcommand{\CN}{{\cal N}}
\newcommand{\CO}{{\cal O}}

\newcommand{\mO}{{\mathsf{O}}}

\def\IZ{{\mathbb Z}}
\def\IR{{\mathbb R}}

\newcommand{\tr}{{\rm Tr}}
\newcommand{\re}{{\rm e}}
\newcommand{\ri}{{\rm i}}
\newcommand{\rd}{{\rm d}}

\newcommand{\Tr}{\mathop{\rm Tr}\nolimits}

\newcommand{\be}{\begin{equation}}
\newcommand{\ee}{\end{equation}}
\newcommand{\ba}{\begin{aligned}}
\newcommand{\ea}{\end{aligned}}
\newcommand{\ben}{\begin{eqnarray}\displaystyle}
\newcommand{\een}{\end{eqnarray}}

\def\({\left(}
\def\){\right)}
\newcommand{\nn}{\nonumber \\}

\newdimen\tableauside\tableauside=1.0ex
\newdimen\tableaurule\tableaurule=0.4pt
\newdimen\tableaustep
\def\phantomhrule#1{\hbox{\vbox to0pt{\hrule height\tableaurule width#1\vss}}}
\def\phantomvrule#1{\vbox{\hbox to0pt{\vrule width\tableaurule height#1\hss}}}
\def\sqr{\vbox{%
  \phantomhrule\tableaustep
  \hbox{\phantomvrule\tableaustep\kern\tableaustep\phantomvrule\tableaustep}%
  \hbox{\vbox{\phantomhrule\tableauside}\kern-\tableaurule}}}
\def\squares#1{\hbox{\count0=#1\noindent\loop\sqr
  \advance\count0 by-1 \ifnum\count0>0\repeat}}
\def\tableau#1{\vcenter{\offinterlineskip
  \tableaustep=\tableauside\advance\tableaustep by-\tableaurule
  \kern\normallineskip\hbox
    {\kern\normallineskip\vbox
      {\gettableau#1 0 }%
     \kern\normallineskip\kern\tableaurule}%
  \kern\normallineskip\kern\tableaurule}}
\def\gettableau#1{\ifnum#1=0\let\next=\null\else
\squares{#1}\let\next=\gettableau\fi\next}

\newcommand{\figref}[1]{Fig.~\protect\ref{#1}}
\title{\boldmath   {Argyres-Douglas theories, Painlev\'e II and quantum mechanics }}

\author{Alba Grassi$^{a}$ and Jie Gu$^{b}$ }
\affiliation{
$^a$Simons Center for Geometry and Physics,\\
 SUNY, Stony Brook, NY, 1194-3636, USA \\
 \\
 $^b$Laboratoire de Physique Th\'eorique de l'\'{E}cole Normale Sup\'erieure\\
 CNRS, PSL Research University, Sorbonne Universit\'{e}s, UPMC, 75005 Paris, France \\}

\emailAdd{agrassi@scgp.stonybrok.edu, gu@lpt.ens.fr}

\abstract{We show in details that the all order genus expansion of the two-cut  Hermitian cubic matrix model  reproduces the perturbative expansion of  the $H_1$ Argyres-Douglas theory coupled to the $\Omega$ background. In the self-dual limit we use the Painlev\'e/gauge  correspondence and we show that, after summing over all instanton sectors, the two-cut cubic matrix model computes the tau function of  Painlev\'e II without taking any double scaling limit or adding any external fields.  We decode such solution within the context of trans-series. Finally in the Nekrasov-Shatashvili limit we connect the $H_1$ and the $H_0$ Argyres-Douglas theories to the quantum mechanical models with cubic and double well potentials.
 }

\begin{document}
\maketitle
\flushbottom

\section{Introduction}

	Argyres-Douglas (AD) theories are four dimensional $\CN=2$ superconformal field theories which were first discovered at special points in the moduli spaces of 4d $\CN=2$ SQCDs where mutually non-local dyons become massless simultaneously \cite{ad1,ad2}. 
	Examples include the $H_0, H_1$, and $H_2$ theories which are limits of $SU(2)$ SQCDs with $N_f=1,2$ and $3$ flavors respectively.
	Matrix model expressions for a number of AD theories have been conjectured in \cite{tc12,Rim:2012tf}. 
	Recently, AD theories have been studied  in connection with the theory of Painlev\'e equations. It was first observed in \cite{Kajiwara:2004ri} that the Seiberg-Witten (SW) curves of four dimensional $\CN=2$ $SU(2)$ SQCDs theories can be extracted from Painlev\'e equations. This observation was made concrete in the breakthrough works of \cite{gil1,gil} which established a precise correspondence between the tau functions of Painlev\'{e} VI, V, III and $\CN=2$ $SU(2)$ SQCDs in the self-dual $\Omega$ background. This picture was recently further generalized in \cite{betal} showing that the partition functions of the $H_0, H_1$ and $H_2$ AD theories compute the tau functions of the Painlev\'e I, II and IV equations.  
	One of the purposes of this paper is to combine  these recent progress in the theory of Painlev\'e equations with some previous works on AD theory, matrix models and resurgence. 
	
	More precisely  in section \ref{sec:mmandad} we show that the all order genus expansion of the $H_1$ theory in the   magnetic frame coupled to the $\Omega$ background is identical to the all order 't Hooft expansion of the $\beta$ deformed cubic matrix model in the two-cut phase. 
	In view of the Painlev\'e/gauge correspondence \cite{betal}, we will focus on the the non-deformed case, i.e.~$\beta=1$, which is defined as 
	\begin{equation}\label{eq:mm}
		Z(N) = \frac{1}{\text{vol}(U(N))} \int \rd \Phi \re^{-\tfrac{1}{g_s}\Tr W(\Phi)} \ ,
	\end{equation}
	where $\Phi$ is an $N\times N$ Hermitian matrix, and the potential is
	\begin{equation}\label{eq:w1i}
		W(x) = \frac{m}{2} x^2 + \frac{1}{3}x^3 \ .
	\end{equation}
	In the one-cut phase of the model all the $N$ eigenvalues of $\Phi$ condensate around the minimum $x = 0$ of the potential and it is possible to show that there exists a double scaling limit of this model 	
	\be
		N \to \infty, \quad t = g_s N \to t_c= \frac{m^3}{12 \sqrt{3}} \ ,
	\ee
	where one reproduces the solution to the Painlev\'e I equation. See \cite{kkn} for a simple derivation and  \cite{DiFrancesco:1993cyw} for a review and a list of references. In the two-cut phase instead one assumes that $N_1$ eigenvalues  condensate around  $x=0$ and $N_2$ eigenvalues around the other critical point of \eqref{eq:w1i} namely $x=-m$. In this case it is possible to write the model as \cite{Cachazo:2001jy} 
	\be \label{eq:mm2i} 
		Z(N_1,N_2)={1\over {\rm Vol}(U(N_1)) \times {\rm Vol}(U(N_2))}\int D\Phi_1 D\Phi_2 {\rm e}^{-{1\over g_s}\left(W_1(\Phi_1)+W_2(\Phi_2)+W_{\rm int}(\Phi_1, \Phi_2)\right)},  
	\ee
	where $W_1 (\Phi_1)$ and  $W_2 (\Phi_2)$ are cubic potentials while $W_{\rm int}(\Phi_1,\Phi_2)$ is an interaction term taking into account the distribution of eigenvalues between the two critical points of the cubic potential (see equation \eqref{eq:potdef} for the precise definition).  
	This matrix model was studied in great details in \cite{Cachazo:2001jy,kmt,kmr}, which we quickly review in section~\ref{sec:mm}, as it describes topological string theory on some particular Dijkgraaf-Vafa geometry.  The refinement of such topological string theory is captured by the $\beta$ deformation of the matrix model \eqref{eq:mm2i}  which has been studied in detail in \cite{huangb} and whose explicit expression is later given in equation \eqref{eq:zn1n2ib}.
	We demonstrate in detail in sections~\ref{sec:H1}, \ref{sec:iden} that such a matrix model also computes the partition function of the $H_1$ theory coupled to the $\Omega$ background where 	\be \beta=-\epsilon_1/ \epsilon_2 \ee 	and we note by	$\epsilon_i$  the $\Omega$ background regulators.

	Thanks to the relation between AD theories in the self-dual background and Painlev\'{e} equations established in \cite{betal}, we can then connect the cubic matrix model \eqref{eq:mm2i}  to the $\tau$ function of the Painlev\'e II equation (PII) at long  time $T\rightarrow \infty$ without taking any double scaling limit, and for generic values of the integration constants as we show in section~\ref{sec:finn}. We would like to note that the two-cut phase of the {\it quartic} matrix model is known to be related to Painlev\'e II in a particular double scaling limit; see for instance \cite{Schiappa:2013opa,mmnp} and references therein. However in this work instead  we consider the two-cut case of the {\it cubic} matrix model and we do not take any scaling limit.
	 The $\tau$-function of PII has the following structure \cite{betal,Its:2016jkt}
	\begin{equation}\label{eq:tauintro}
		\tau_{II}(T) \propto \sum_{n\in \IZ} { \re^{\ri n \rho}}\re^{\ri n (2\sqrt{2}/3)T^{3/2}} \CG'(T,\nu+n) \ ,\quad T\rightarrow \infty \ ,
	\end{equation}	
	where $(\nu, \rho)$ are integration constants and $T$ is the time. In  \cite{betal} the quantity $ \CG'$ is given  as a series expansion in $T^{-3/2}$ and  the first few terms have been computed explicitly.
	As explained in section~\ref{sec:finn} we find that the two-cut cubic matrix model is identified with the summand $\CG'(T,\nu)$. This observation enables us to compute  $\CG'(T,\nu)$ at large $T$ up to very high orders and then to study the convergence properties of the solution proposed in \cite{betal,Its:2016jkt}.  For some particular choice of integration constants  we can give an all order formula for  $\CG'(T,\nu)$ (see equations  \eqref{eq:M-limit-H1},\eqref{eq:ds}). We find that the long time expansion of \cite{betal,Its:2016jkt} is in fact divergent and we argue in section~\ref{sec:pa} that the summation over $n$ in \eqref{eq:tauintro} amounts to a sum over all instanton sectors in the matrix model, and that the instantons have the correct action as extracted from the analysis of the large order behaviour of  $\CG'(T,\nu)$,  namely 
	\be 
		\ri \, 2\sqrt{2}/3  \ .
	\ee
	At the end of section~\ref{sec:pa}  we briefly discuss Borel summability of \eqref{eq:tauintro}.

	Finally in section \ref{sec:qm} we discuss the $H_1$ and $H_0$ theories in the NS phase of the $\Omega$ background and we show that these theories can be used to compute the all order WKB periods of the QM models with the cubic and the double well potentials. This provides a gauge theory justification for the holomorphic anomaly algorithm proposed in \cite{csm} to determine the spectra of these QM models.
	
\section{Matrix model and Argyres-Douglas theory}\label{sec:mmandad}

	In this section, we show that perturbatively the two-cut phase of the $\beta$ deformed  Hermitian cubic matrix model can be identified with the $H_1$ Argyres-Douglas theory in the magnetic frame coupled to the $\Omega$ background. The 't Hooft expansion of the Hermitian cubic matrix has been discussed for instance in \cite{Cachazo:2001jy,kmt,kmr,Dijkgraaf:2002pp}, while its $\beta$ deformation was studied in \cite{huangb}, see also \cite{Maruyoshi:2014eja,bmt} for more details on the  $\beta$  deformed models.
	Some results for the free energies of the $H_1$ theory can be found for instance in {\cite{Masuda:1996xj,betal}}. We quickly review these two theories, and then demonstrate how they can be identified.  Our derivation is rigorous for the case $\beta=1$ but it relies on some conjectural results of \cite{huangb} for the case $\beta \neq 1$.
	
	Physically, one can argue in favour of a connection between this matrix model and the $H_1$ theory by following \cite{tc12,Rim:2012tf} even though  the details of the connection and the precise dictionary between these two theories was not spelled out in these references.
	The proposal of \cite{tc12,Rim:2012tf} was intended to give matrix model realisations for the irregular conformal blocks studied in \cite{Bonelli:2011aa,Gaiotto:2012sf} and it involves in general  a Riemann sphere with an irregular singularity at infinity and a regular singularity at $z=0$. This gives a matrix model with potential 
	\be 
		{1\over g_s}V(z)=\alpha^{(0)} \log z - \sum_{k=1}^n{c_k^{(\infty)}z^k\over k}. 
	\ee
	The coefficient $\alpha^{(0)}$ characterises the regular singularity at $z= 0$. 
	Then one can argue that by taking the limit $\alpha^{(0)} \to 0$ one removes the regular singularity, in which case one arrives at the $A_{2n-3}$ AD theories (only irregular singularity at infinity), to which the $H_1$ theory belongs. This is how one can argue for  a connection between the $H_1$ theory and the cubic matrix model from the perspective of \cite{tc12}. Notice however that  strictly speaking in the approach of \cite{tc12} removing all the regular singularities may not be completely justified\footnote{We would like to thank Giulio Bonelli, Kazunobu Maruyoshi and Alessandro Tanzini raising this issue as well as for clarifications and discussions on these models.},  and furthermore the details of the connection and a precise dictionary was not proposed.
	On the other hand, even if the physical justification given above is not very rigorous, in the forthcoming section, after establishing the precise dictionary, we will verify by a direct computation that the model \eqref{eq:mm2i} computes the all order genus expansion of the $H_1$ theory.

 %We note that a  connection between this matrix model and the $H_1$ theory was already conjectured in \cite{tc12} even though the details of the connection and the precise dictionary between these two theories was not spelled out in \cite{tc12}. In the forthcoming section we will establish the precise dictionary and verify perurbatively the conjceture to all order in genus expansion. 

	\subsection{The two cut phase of the cubic model }\label{sec:mm}
	
	We first study in some details the case $\beta=1$ since we will need it in the forthcoming section and we briefly illustrate the case of generic $\beta$ at the end of this subsection. 
	
	We consider the hermitian matrix model 
	\begin{equation}\label{eq:mm}
		Z(N) = \frac{1}{\text{vol}(U(N))} \int \rd M \re^{-\tfrac{1}{g_s}\Tr W(M)} \ ,
	\end{equation}
	where $M$ is an $N\times N$ hermitian matrix, and the potential is
	\begin{equation}
		W(x) = \frac{m}{2} x^2 + \frac{1}{3}x^3 \ .
	\end{equation}
	This potential has two critical points at $x=0$ and $x = -m$ respectively. Let us consider the vacuum where $N_1$ eigenvalues of $M$ condensate at the critical point $x=0$, while $N_2$ eigenvalues condensate at $x = -m$, such that $N_1 +N_2 = N$. When expanded around this vacuum, the matrix model can be written as
%Then one has
	\be \label{eq:zn1n2i}
		Z(N_1, N_2)= {1\over N_1! N_2! } \int \prod_{k=1}^N \frac{\rd x_k}{2\pi} \,{\overline {\Delta}}^2(x)\re^{- \tfrac{1}{g_s}  {\overline W}(x_k)} \ ,
	\ee
	where 
	 \be \label{eq:deltaa}
		 \overline \Delta^2(x)=\prod_{1 \leq i_1< i_2 \leq N_1}(x_{i_1}-x_{i_2})^2\prod_{N_1+1 \leq j_1< j_2 \leq N}(x_{j_1}-x_{j_2})^2\prod_{  1 \leq i \leq N_1, N_1+1 \leq j \leq N } (x_{i}-x_{j}+m)^2 \ , 
	\ee 
	and 
	\be \label{eq:pott}
		{\overline W}(x_k) = \sum_{i=1}^{N_1}\left( \frac{m}{2} x_i^2 + \frac{1}{3}x_i^3\right) +\sum_{i=N_1+1}^{N}\left(- \frac{m}{2} x_i^2 + \frac{1}{3}x_i^3\right) +{m^3\over 6 } N_2 \ . 
	\ee
	In \eqref{eq:zn1n2i} we have multiplied \eqref{eq:mm} by 
	\be {N! \over N_1! N_2!} \ee
	to take into account the different ways in which the eigenvalues distribute.  We can view \eqref{eq:zn1n2i} as a two-matrix model integral \cite{kmt, kmr,Cachazo:2001jy,Dijkgraaf:2002pp} 
	\be \label{eq:mm2} 
		Z(N_1,N_2)={1\over {\rm Vol}(U(N_1)) \times {\rm Vol}(U(N_2))} \int \rd\Phi_1 \rd\Phi_2 \,{\rm e}^{-{1\over g_s}\left(W_1(\Phi_1)+W_2(\Phi_2)	+W(\Phi_1, \Phi_2)\right)},  
	\ee
	where $\Phi_1, \Phi_2$ are $N_1\times N_1$ and $N_2\times N_2$ matrices respectively, and the potentials are
	\be \label{eq:potdef}\ba
		W_1 (\Phi_1)=&\tr\Bigl({1\over 2} m \Phi_1^2 + {1 \over 3} \Phi_1^3\Bigr)\ , \\
		W_2 (\Phi_2)=&-\tr\Bigl({1\over 2} m \Phi_2^2 - {1 \over 3} \Phi_2^3\Bigr)\ , \\
		W_{\rm int}(\Phi_1,\Phi_2)=&N_2 {m^3\over 6}+2 N_1 N_2\ln\left(m\right)+ 2\sum_{k=1}^{\infty} \frac{1}{k m^k} \sum_{p=0}^k (-1)^p{k\choose p}{\rm tr}\,\Phi_1^p~ {\rm tr}\,\Phi_2^{k-p} \ .
	\ea \ee
	Note that for the above two-matrix model to be perturbatively well defined we have to choose
	$\Phi_1$ to be hermitian and $\Phi_2$ anti-hermitian. In other words, in the eigenvalue formalism \eqref{eq:zn1n2i}, we choose
	\be
		 x_i=\left\{ \begin{array} {cc}
		 &\in \IR \quad \quad i\leq N_1 \\
		  & \in \ri \IR \quad \text{otherwise}
		 \end{array}\right. \ .
	\ee
	In this section we are interested in the 't Hooft expansion of the matrix model  \eqref{eq:zn1n2i}
	\be \label{eq:thooft}
		N_i \rightarrow \infty \ ,\; g_s \rightarrow 0\ , \quad S_i=g_s N_i  \;\text{fixed} \ ,\quad i=1, 2\ . 
	\ee
	We have defined the partial 't Hooft parameters $S_{1,2}$ such that
	\begin{equation}
		S_1 + S_2 = t = g_s N \ .
	\end{equation}
	In this regime the matrix model integral can be canonically expanded as
	\be \label{eq:zmme}
		\log Z(N_1,N_2)=\sum_{g\geq0}g_s^{2g-2}F_g(S_1, S_2) \ ,
	\ee
	where $F_g$ are the genus $g$ free energies.
	
	The free energies of the matrix model can be computed from the spectral curve of the Hermitian matrix model and the associated 1-differential. The spectral curve reads
	\begin{equation}\label{eq:mmsc}
		\mathcal{C}^{\text{mm}}: \; y^2 = W'(x)^2 + f(x) =x^2(x+m)^2 + \lambda x + \mu \ ,
	\end{equation}
	while the associated the 1-differential is
	\be\label{eq:1formmm}
		\Omega^{\text{mm}} = y(x) \rd x \ . 
	\ee
	The spectral curve is defined to be the deformation of the singular curve
	\begin{equation}
		y^2 = W'(x)^2 \ ,
	\end{equation}
	where the two singular points $a_1, a_2$ ($a_2>a_1>0$) are the two critical points of the matrix model potential. After turning on the deformation $f(x) = \lambda x + \mu$, the two singular points extend to two branch cuts on the real axis, whose endpoints we denote by $a_1^-, a_1^+$ and $a_2^-, a_2^+$ respectively. Sometimes we denote the branch points also by 
	\begin{equation}
		(a_1^-, a_1^+, a_2^-, a_2^+) \;\rightarrow\; (x_1, x_2, x_3, x_4) \ ,
	\end{equation}
	from leftmost to rightmost, and the spectral curve then reads
	\be  
		y^2 =\prod_{i=1}^4 (x-x_i) \ .
	\ee
	Let us introduce the variables 
	\begin{equation}
		z_1 = \frac{1}{4}(x_2 -x_1)^2 \ ,\quad z_2 = \frac{1}{4}(x_4 - x_3)^2 \ ,
	\end{equation}
	that measure the width of the branch cuts. Clearly the spectral curve is singular when $z_1 = 0$ or $z_2 = 0$. The spectral curve can also become singular when $x_3 \rightarrow x_2$ so that the two branch cuts fuse into one. This is what is called the dual conifold point in the language of \cite{csm}.
	
\begin{figure}
		\centering
		\includegraphics[width=0.6\linewidth]{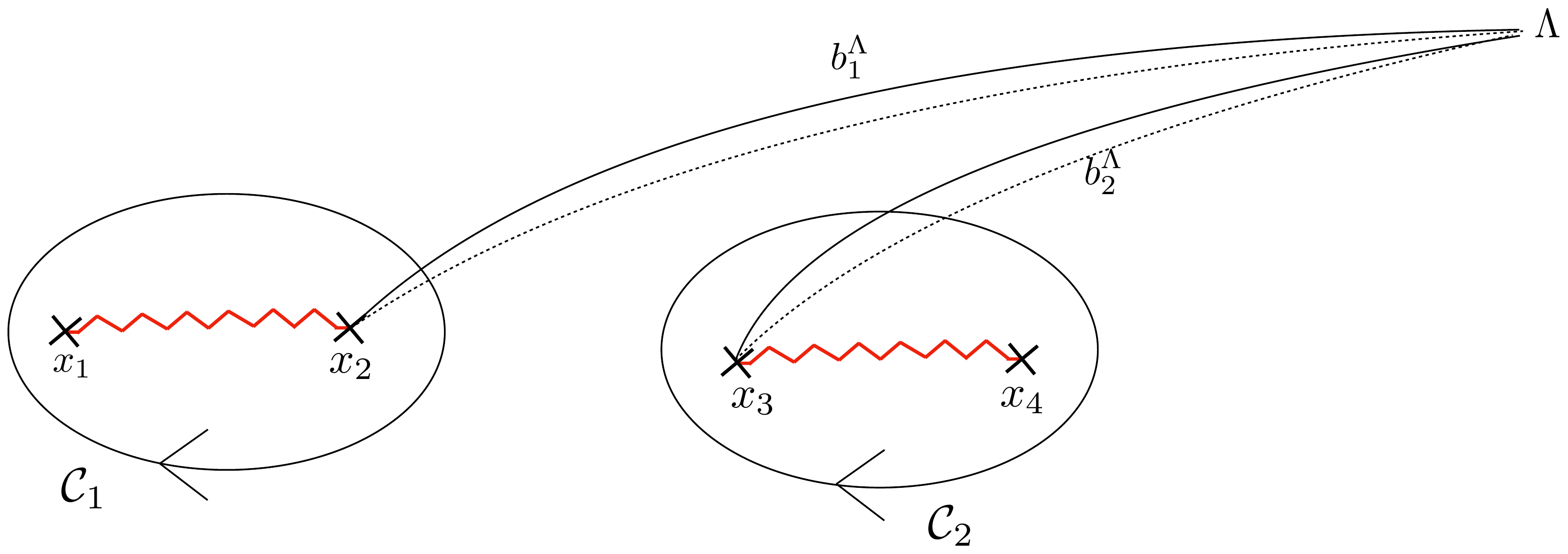}
		\caption{Path of integrals in the period calculations \cite{kmr}.}\label{fg:periods}
	\end{figure}

	We are interested in the limits when $z_{1,2}$ are small, where the partial 't Hooft parameters $S_{1,2}$ are locally good coordinates on the moduli space. They can be identified with the integrals of the canonical 1-form along the cycles $C_{1,2}$ that surround the branch cuts (see \figref{fg:periods})
	\begin{equation}\label{eq:periods}
		S_i = \frac{1}{2\pi \ri}\int_{a_i^-}^{a_i^+} \Omega^{\text{mm}} \ ,\quad i=1,2\ .
	\end{equation}
	Performing the period integrals explicitly, one finds \cite{Cachazo:2001jy}
	\be \label{eq:Sian}
	\ba
		& S_1 = \frac{1}{4} z_1 \cdot I - \frac{1}{2I} K(z_1, z_2, I^2)\ ,\\
		& S_2 = -\frac{1}{4}z_2 \cdot I + \frac{1}{2I} K(z_1, z_2, I^2) \ .
	\ea
	\ee
	Here $I$ is
	\begin{equation}
		I^2 = \frac{1}{4}((x_3 + x_4)-(x_1 + x_2))^2 = m^2 - 2(z_1 + z_2) \ ,
	\end{equation}
	and $K(z_1, z_2, I^2)$ is a transcendental function symmetric in $z_1, z_2$, whose expansion reads\footnote{When $n+m=0$, the product $(n+m)\, \Gamma(2n+2m)$ in the numerator takes the limit value which is 1/2.}
	\begin{equation}
		K(z_1,z_2,I^2) = \sum_{n,m=0}^\infty \frac{(n+m)\, \Gamma(2n+2m)}{2^{2n+2m+1}\Gamma(n+2)\Gamma(n+1)\Gamma(m+2)\Gamma(m+1)}\frac{z_1^{n+1}z_2^{m+1}}{(I^2)^{m+n}} \ .
	\end{equation}
	 Let us introduce
	\begin{equation}\label{eq:st}
	\begin{aligned}
		& t = S_1 + S_2  = \frac{1}{4}(z_1 - z_2)\cdot I\ ,\\
		& s = S_1 - S_2 \ . 
	\end{aligned}
	\end{equation}
	Clearly the period $t$ is only an algebraic function of $z_1, z_2$, and borrowing the terminology of gauge theories it can be termed a mass parameter, while $s$, being a transcendental function, is the only true modulus. In terms of the parameters $\lambda,\mu$ of the spectral curve that control the width of the branch cuts, since
	\begin{equation}\label{eq:t-lambda}
		t = S_1 + S_2 = -\frac{1}{2}\oint_\infty \frac{\rd x}{2\pi \ri}y(x) = \frac{\lambda}{4} \ ,
	\end{equation}
	we conclude that $\lambda$ is associated to the mass parameter, while $\mu$ is associated to the true modulus $s$.

	In order to compute the planar free energy $F_0$, we introduce the dual periods $\Pi_i$, which are integrals of the 1-differential along the dual cycles $b_{1,2}^\Lambda$ that extend from $a_1^+, a_2^-$ to the UV regulation point $\Lambda$ (see Fig.~\ref{fg:periods})
	\begin{equation}
		\Pi_{i} = \int_{b_i^\Lambda} \Omega^{\text{mm}} \ , \quad i=1,2 \ .
	\end{equation}
	The planar free energy in the small $z_{1,2}$ limit is determined by the special geometry relation, which reads, 
	\begin{equation}\label{eq:spec-geom}
		\frac{\partial F_0}{\partial S_i} = \Pi_i \ ,\quad i =1,2 \ ,
	\end{equation}
	or if only the true modulus $s$ is used
	\begin{equation}\label{eq:F0mm}
		\frac{\partial}{\partial s}F_0 = \frac{1}{2}(\Pi_1 - \Pi_2) \ .
	\end{equation}
	In addition, since
	\begin{equation}
		\frac{\partial}{\partial \mu} \Omega^{\text{mm}} = {1\over 2}\frac{\rd x}{y(x)} \ ,
	\end{equation}	
	the periods $S_{1,2}, \Pi_{1,2}$ can be expressed in terms of elliptic integrals
	\begin{equation}\label{eq:mmdS}
	\begin{aligned}
		2\pi\ri \frac{\partial}{\partial \mu} S_1 &= -2\pi\ri \frac{\partial}{\partial \mu} S_2 ={1\over 2} \int_{x_1}^{x_2} \frac{\rd x}{y(x)} = -\frac{\ri}{\sqrt{(x_1 - x_3)(x_2 - x_4)}}K(k^2) \ , \\
		\frac{\partial}{\partial \mu}(\Pi_1 - \Pi_2) &= {1\over 2}\int_{x_2}^{x_3} \frac{\rd x}{y(x)} = \frac{1}{\sqrt{(x_1 - x_3)(x_2 - x_4)}} K(k'^2) \ ,
	\end{aligned}
	\end{equation}
	with
	\begin{equation}
		k^2 = \frac{(x_1 - x_2)(x_3 - x_4)}{(x_1 - x_3)(x_2 - x_4)} \ ,\quad k'^2 = 1-k^2 \ .
	\end{equation}
	As a result, we can also write the planar free energy as
	\begin{equation}
		\partial^2_s F_0 = -\frac{\pi}{2}\frac{K(k'^2)}{K(k^2)} \ .
	\end{equation}

The genus one free energy $F_1$ can be computed using the Akemann formula for a two-cut matrix model \cite{Akemann:1996zr,kmr}
	\be \label{eq:f1a} F_1 = -\frac{1}{24}\sum_{i=1}^4 \log M_i^{(1)} - \frac{1}{2}\log K(k^2) -\frac{1}{12}\log \Delta + \frac{1}{8}\log(x_1 -x_3)^2 + \frac{1}{8}(x_2 -x_4)^2 + \text{const.} \ , \ee
	where $\Delta$ is the discriminant
	\begin{equation} \label{eq:dis}
		\Delta = \prod_{i<j} (x_i - x_j)^2 \ ,
	\end{equation}
	and the moments $M_i^{(1)}$ are all $1$ in the cubic matrix model. Explicitly when $s,t$ are small, we have \cite{kmt} 
	\begin{equation}\label{eq:mmF1}
		F_1 = -\frac{1}{12}\log (s^2 - t^2) +\frac{1}{6m^3} s + \frac{1}{12m^6}(45s^2 - 17t^2)+ \frac{1}{36m^9}(3101 s^3 - 1773 s t^2) + \mO(s^3, t^3) \ .
	\end{equation}
	Note that by applying \eqref{eq:mmdS} the Akemann formula can be cast in the form (up to a constant)	\begin{equation}\label{eq:F1mm}
		F_1  =- \frac{1}{2}\log \left( \frac{\partial s}{\partial \mu}\right) - \frac{1}{12}\log \Delta \ .
	\end{equation}
	
	Finally, the free energies of higher genera can be computed by using the holomorphic anomaly equations \cite{bcov} as demonstrated in \cite{kmr}. It requires as initial data the flat coordinates, $s$ and $t$, the planar and genus one free energies in the small $z_{1,2}$ limit, as well as the transformation of these local data to the vicinity of the conifold singularity. In turns these quantities are determined by the spectral curve and the the choice of  1-differential. The higher genera free energies of the hermitian matrix model could also be computed by the means of topological recursion \cite{eo}, using the spectral curve \eqref{eq:mmsc} and the 1-differential \eqref{eq:1formmm}.
	
	Let us now briefly discuss the $\beta$ deformation of the cubic model namely \cite{huangb}
		\be \label{eq:zn1n2ib}
		Z_{\beta}(N_1, N_2)= {1\over N_1! N_2! } \int \prod_{k=1}^N \frac{\rd x_k}{2\pi} \,{\overline {\Delta}}^{2\beta}( x)\re^{- \tfrac{\beta}{g_s}  {\overline W}(x_k)} \ .
	\ee
	In the t' Hooft regime \eqref{eq:thooft} one has
	\be \label{eq:zmmeb}
		\log Z_{\beta}(N_1,N_2)=\sum_{g,n\geq0}g_s^{2(g+n)-2}(-\beta)^{1-g-2n}(\beta-1)^{2n}F_{g,n}(S_1, S_2) \ .
	\ee
	When $\beta=1$ we recover  \eqref{eq:zmme} with the identification 
	\be F_{g,0}=F_{g}.\ee
	It was conjectured and tested in \cite{huangb} that  $F_{g,n}$ can be computed recursively by solving the {\it refined} holomorphic anomaly equations \cite{hk,kwal}. The latter are extension of the holomorphic anomaly equations \cite{bcov} and their solution requires an additional piece of initial condition, namely the knowledge  of  $F_{0,1}$: the  genus one free energy in the NS limit. For the matrix model \eqref{eq:zn1n2ib} it was conjectured and tested in \cite{huangb} that $F_{0,1}$ is given by
	\be\label{eq:f1n} 
		F_{0,1}= - \frac{1}{24}\log \Delta\ ,
	\ee
	where $\Delta$ is the discriminant \eqref{eq:dis}. To our knowledge a rigorous derivation of this statement in matrix models is missing.

	\subsection{The $H_1$ Argyres-Douglas  theory}\label{sec:H1}

	We quickly review the computational aspect of the Argyres-Douglas theory called $H_1$ which lies inside the moduli space of $\CN=2$ $SU(2)$ SQCD with two flavour $N_f=2$ \cite{ad2}. Just like the Seiberg-Witten theory, the $H_1$ theory is completely encoded in the spectral curve which is given by (we follow the notation of \cite{betal})
	\be\label{eq:ch1} 
		\mathcal{C}^{H_1}: \ y_{\rm H_1}^2 = x^4 + 4c x^2 + 2{m_{\rm H_1}} x + {u} \ ,
	\ee
	as well as the associated canonical one-form
	\be\label{eq:1fh1} 
		\Omega_{H_1} = y_{H_1}(x) \rd x \ .
	\ee
	The parameter $m_{\rm H_1}$ is the mass parameter, ${c}$ is the deformation parameter away from the conformal point, while ${u}$ is the Coulomb modulus. Let $e_i$ be the four roots of \eqref{eq:ch1}. The spectral curve can be written as 
	\be 
		y_{\rm H_1}^2=\prod_{i=1}^4(x-e_i) \ ,
	\ee
	and we introduce the periods of the 1-form (we follow the notation of \cite{Masuda:1996xj})
	\be \label{eq:period} 
		a=\int_{e_2}^{e_3} y_{\rm H_1}(x) \rd x \ ,\quad  \quad a_D= \frac{1}{2\pi\ri}\int_{e_1}^{e_3} y_{\rm H_1}(x) \rd x \ .
	\ee

	On the other hand, the Coulomb branch is also parameterized by the mass parameter $m_{\rm H_1}$, which is given by the residue of the canonical 1-form at the infinity of the $x$-plane. We choose to treat $m_{\rm H_1}$ in a symmetric way. For this purpose, we introduce 
	\be 
		{\overline a_D}= {1\over 2 \pi \ri }\int_{e_4}^{e_2} y_{\rm H_1} (x) \rd x   \ ,
	\ee
	which satisfies
	\be   
		{a}_D - \overline{a}_D= m_{\rm H_1}/2\ .
	\ee
	It is then also convenient to introduce
	\begin{equation}
		\tilde{a}_D = a_D + \overline{a}_D = 2a_D - m_{\rm H_1}/2 \ .
	\end{equation}
	In particular in the massless limit $m_{\rm H_1} = 0$, one finds that
	\begin{equation}
		a_D = \overline{a}_D = \tilde{a}_D/2 \ .
	\end{equation}
		
	The singularities of the Coulomb branch are given by the zeros of the discriminant of the spectral curve
	\be  
		\Delta = 256 u^3 - 2048c^2 u^2 + 4096 c^4 u + 2304 m_{\rm H_1}^2 c u -432 m_{\rm H_1}^4 - 1024 m_{\rm H_1}^2 c^3 \ . 
	\ee
	One reads off three singular points (perturbatively in small $m_{\rm H_1}$)
	\be\label{eq:3pts}
	\ba
		u^{(1)} &= \frac{m_{\rm H_1}^2}{4c} + \mO(m_{\rm H_1}^4) \ ,\\
		u^{(2)} &= 4c^2 + 2\sqrt{2}\ri m_{\rm H_1}c^{1/2} -\frac{m_{\rm H_1}^2}{8c} + \frac{\ri m_{\rm H_1}^3}{64\sqrt{2} c^{5/2}} + \mO(m_{\rm H_1}^4) \ ,\\
		u^{(3)} &= 4c^2 - 2\sqrt{2}\ri m_{\rm H_1}c^{1/2} -\frac{m_{\rm H_1}^2}{8c} - \frac{\ri m_{\rm H_1}^3}{64\sqrt{2} c^{5/2}} + \mO(m_{\rm H_1}^4) \ .
	\ea
	\ee
	 We refer to $u^{(1)} $ as the electric point while $u^{(2)} $ and $u^{(3)}$ correspond to the magnetic and dyonic points.
	 		
	In this section we focus on the magnetic frame, and consider the $H_1$  theory coupled to the $\Omega$ background  \cite{n,Nekrasov:2003rj} where the two $\Omega$ regulators are 
	\be 
		\epsilon_1, ~\epsilon_2 \ . 
	\ee
	In the magnetic frame, the good local coordinate is the period $a_D$, and the partition function enjoys the genus expansion
	\be\label{eq:zad} 
		\log Z^{\rm D}(a_D)=F^{\rm D}(a_D)=\sum_{g,n\geq 0} F_{g,n}^{(D)}(a_D) (\epsilon_1 \epsilon_2)^{g-1} (\epsilon_1+\epsilon_2)^{2 n}\ .
	\ee	
	When $\epsilon_1=-\epsilon_2=\epsilon$ we note 
	\be 
		F_{g,0}^{(D)}=  F_{g}^{(D)}\ . 
	\ee
	The prepotential $F_0^{(D)}$ in the magnetic frame is then determined by the following special geometry relation
	\begin{equation}\label{eq:f0H1}
		\frac{\partial}{\partial \tilde{a}_D} F_0^{(D)} = \frac{a}{2} \ .
	\end{equation}
	The genus one free energies of the gauge theory are given by \cite{Huang:2006si,Huang:2009md}
	\be \label{eq:f1H1n} 
		F^{(D)}_1 = -\frac{1}{2}\log\left(\frac{\rd \tilde a_D}{\rd u}\right) - \frac{1}{12}\log \Delta \ , 
	\ee
	and
	\be \label{eq:f1H1} 
		F^{(D)}_{0,1} =- \frac{1}{24}\log \Delta \ , 
	\ee
		Finally the higher genus free energies $ F_{g,n}^{(D)}$ can also be determined by using the refined holomorphic anomaly equations \cite{Huang:2006si,Huang:2009md, hk,kwal}.

	\subsection{Identifying gauge theory with matrix model}\label{sec:iden}

	In this section we show that the all order 't Hooft expansion of the matrix model \eqref{eq:zmmeb}  is identical to the all order genus expansion of the AD theory $H_1$ \eqref{eq:zad} in the $\Omega$ background.
	To start with we would like to identify the spectral curves and the choices of canonical one form. After taking the shift of variables
	\be\label{eq:shift}
		x \mapsto x-m/2 \ ,
	\ee
	the spectral curve \eqref{eq:mmsc} of the matrix model becomes
	\be\label{eq:c2}
		y^2 =x^4 - \frac{m^2}{2}x^2 + \lambda x + \left(\frac{m^4}{16} - \frac{\lambda m}{2} + \mu\right) \ ,
	\ee
	while the associated 1-differential remains the same. It is then easy to see that both the spectral curve and the canonical differential of the cubic matrix model can be identified with those of the $H_1$ theory, i.e.~\eqref{eq:ch1} and \eqref{eq:1fh1}, provided we use the following dictionary
	\begin{equation}\label{eq:dict}
	\begin{aligned}
		c &= -\frac{m^2}{8} \ ,\\
		m_{\rm H_1} &= \frac{\lambda}{2}=2 t\ ,\\
		u &= \frac{m^4}{16} - \frac{\lambda m}{2} + \mu \ .
	\end{aligned}
	\end{equation}
	In particular, the mass parameter ${m_{\rm H_1}}$ and the Coulomb modulus ${u}$ of the $H_1$ theory are identified correspondingly with the mass parameter $\lambda \propto t$ and the true modulus $\mu$ (up to a shift) of the matrix model. Therefore, the Coulomb branch of the $H_1$ theory can be identified with the complexified moduli space of the cubic matrix model: both of them have the same singular points and the same metric. 

	Let us make the identification of singularities more explicit. We first demonstrate that the singularity of the matrix model $z_1\rightarrow 0$ or $z_2 \rightarrow 0$ should be identified with $u^{(2)}$ or $u^{(3)}$ singularities of the $H_1$ theory. Indeed, the two singularities of the $H_1$ theory are related to each other by mapping
	\be 
		m_{\rm H_1} \rightarrow -m_{\rm H_1}
	\ee
	while keeping ${c}$ and ${u}$ fixed. On the matrix model side, if we send $t\to -t$ and keep the combinations $m^2$ and $m^4 - 8\lambda m +16\mu$ unchanged (c.f.\ \eqref{eq:dict}), the four branch points are mapped to
	\begin{equation}
		(x_1, x_2, x_3,x_4) \longmapsto (-x_4, -x_3, -x_2, -x_1) \ ,
	\end{equation}
	which means that we merely exchange $z_1$ and $z_2$. Furthermore, let us take the simple limit $m_{\rm H_1} = 0$, and consider the domain
	\begin{equation}
		{c} < 0 \ ,\quad 4 {c}^2 - u > 0 \ ,
	\end{equation}
	where the four branch points of the spectral curve of the $H_1$ theory lie on the real axis. It is easy then to compute the position of the four branch points and check that the confluent singularity $u^{(2)} = u^{(3)}$ of the $H_1$ theory corresponds to the limit $z_1 = z_2 = 0$ of the matrix model. In addition, a simple calculation in this limit shows that the $u^{(1)}$ singularity of the $H_1$ theory corresponds to the dual conifold point of the matrix model, where $x_2 = x_3$, and two branch cuts fuse into one. To summarize, we have the following correspondence of singular points
	\begin{equation}
		\begin{array}{rcl}
			z_1 = 0 & \longleftrightarrow & u^{(2)} \\
			z_2 = 0 & \longleftrightarrow & u^{(3)} \\
			\text{dual conifold} & \longleftrightarrow & u^{(1)} \ ,
		\end{array}
	\end{equation}
	where we borrow the terminology of \cite{csm} to denote the $u^{(1)}$ singularity.
	In particular the small branch cut limit $S_1\rightarrow 0$ indeed corresponds to the magnetic point where $a_D$ is small.

	Given that the spectral curve and the one form in the two theories are identified, it follows that also the periods coincide. It is  straightforward  to check for instance 	that the following dictionary can be established
	\be 
	\ba
	S_1=  a_D\\
	\ea
	\ee
	provided we identify
	\begin{equation}
		(e_1, e_2, e_3, e_4) \leftrightarrow (\tilde{x_1}, \tilde{x}_3, \tilde{x}_2, \tilde{x}_4) \ ,
	\end{equation}
	where we  denote the four branch points of the spectral curve after the shift \eqref{eq:shift} by $\tilde{x}_i$ ($i=1,2,3,4$). More precisely we identify
	\begin{equation}\label{eq:dict-2}
	\begin{gathered}
		a_D= S_1 \ , \quad \overline{ a}_D=-S_2\ ,\\
		\tilde{a}_D = s \ ,\\
		a = \Pi_1 - \Pi_2 \ .
	\end{gathered}
	\end{equation}
	Consequently, in the magnetic (dyon) frame of the $H_1$ theory and the small $z_1,z_2$ limit of the matrix model, we could identified the planar and genus one free energies of the two theories by comparing \eqref{eq:f0H1}, \eqref{eq:f1H1}, \eqref{eq:f1H1n} and \eqref{eq:F0mm}, \eqref{eq:F1mm},\eqref{eq:f1n}. Higher genera free energies can also be identified since they can be computed by using the refined holomorphic anomaly equations on both sides. 
	
	In fact, at least for the case $\beta=1$, as long as we make the conjecture that the $H_1$ theory has an underlying hermitian matrix model so that the topological recursion \cite{eo} is applicable, the identification of spectral curve and canonical 1-form with the cubic matrix model through \eqref{eq:dict} already suffices to guarantee the all genus expansion of the partition functions of the two theories are in agreement.

	A final note is that the spectral curve of the two-cut solution to the cubic matrix model in the $S_1 = -S_2$ slice itself can be identified with many other theories, like the pure $SU(2)$. 
	 But since their 1-forms are different, not all their free energies can be identified. The case of pure $SU(2)$ is discussed in { \cite{kmt, kmr}}, where it is pointed out that one only has the agreement of $ \partial^2F_0$ and $F_1$.

	\section{The two-cut model and the Painlev\'e/gauge correspondence }\label{sec:finn}
	
	We provide here a concrete link between the two-cut matrix model \eqref{eq:mm2} and the proposal of \cite{betal} where the partition function of the $H_1$ theory was computed in the large $c$ regime  \eqref{eq:ch1}.
	
	We consider the limit
	\begin{equation}\label{eq:M-limit}
		g_s \rightarrow 0 \ ,\quad N_1, N_2\;\; \text{finite}  \ ,
	\end{equation}
	in which case, the matrix integral has the decomposition
	\begin{equation}
		Z_\text{mm}(N_1,N_2) = Z_\text{mm}^{\text{np}}(N_1,N_2) \(1 + \sum_{k=1}^\infty g_s^k Z_\text{mm}^{(k)}(N_1,N_2) \) \ .
	\end{equation}
	In order to see how the nonperturbative contributions $Z_\text{mm}^{\text{np}}(N_1,N_2)$ and perturbative contributions $Z_\text{mm}^{(k)}(N_1,N_2)$ are related to the free energies $F_g(t)$ of the 't Hooft expansion, notice that the latter have the following asymptotic behavior
	\begin{equation}\label{eq:M-limit-mm}
	\begin{aligned}
		F_0(t) &= \frac{t^2}{2}(\log t-1/2) + c^{(0)}_1 t + c^{(0)}_2 t^2 + \CO(t^3) \\
		F_1(t) &= -\frac{1}{12}\log t + c^{(1)}_1 t + \CO(t^2) \ ,\\
		F_g(t) &= \frac{B_{2g}}{2g(2g-2)t^{2g-2}} + c^{(g)}_1 t + \CO(t^2) \ , \quad g\geq 2 \ ,
	\end{aligned}
	\end{equation}
	where for simplicity we take a one-cut matrix model as an example. Plug in $t = g_s N$ and take the limit \eqref{eq:M-limit}, one finds
	\begin{align}\label{eq:z1c}
		& Z_\text{mm}(N) = \exp\(\sum_{g=0}^{\infty} g_s^{2g-2} F_g(t)\) \nn
		= &\exp\( \frac{N^2}{2}\(\log (g_s N) - 1/2\) + \frac{c^{(0)}_1 N}{g_s} + c^{(0)}_2 N^2 - \frac{1}{12}\log (g_s N)+\sum_{g=2}^\infty \frac{B_{2g}}{2g(2g-2) N^{2g-2}}\)\times\nn
		&\exp\(\sum_{g=0}^\infty \sideset{}{'}\sum_{n=1}^{\infty}  g_s^{2g-2+n} c^{(g)}_n N^n\) \ ,
	\end{align}
	where in the last line $\sideset{}{'}\sum$ means $n$ starts from 3 if $g=0$ so that one always has $2g-2+n \geq 1$. Obviously, the second line  in \eqref{eq:z1c}  is $Z_\text{mm}^{\text{np}}(N)$, and other than the ambiguous contributions\footnote{When one computes the planar free energy $F_0(t)$ from the special geometry relation, for instance \eqref{eq:spec-geom}, the linear and quadratic terms are ambiguous.} $c^{(0)}_1, c^{(0)}_2$ this term is universal. The third line expands to $Z_\text{mm}^{(k)}(N)$, and they receive leading contributions of higher genera free energies $F_g(t)$. Therefore the limit \eqref{eq:M-limit} gives us a means to compare more directly higher genera free energies of the two theories.
	
	Let us come back to the two-cut solution to the cubic matrix model. The perturbative contributions have been computed in \cite{kmt}\footnote{There is a little typo in the perturbation contributions computed in equation (4.9) of \cite{kmt}. The 5th order proportional to $g^{10}/m^{15}$ should start with $(9152/5N_1^7-\ldots$ instead of $9152/5(N_1^7-\ldots$.} up to order 6. The first few orders are \cite{kmt}
	\begin{align}  
		 Z_\text{mm}^{(1)}(N_1,N_2)= &\frac{1}{6m^3}\(2(2N_1^3 -15N_1^2N_2 + 15N_1 N_2^2 - 2N_2^3) + (N_1-N_2)\) \ ,\\
		 Z_\text{mm}^{(2)}(N_1,N_2) = & \frac{1}{3m^6}\((8N_1^4-91N_1^3N_2+59N_1^2N_2^2-91N_1N_2^3+8N_2^4)+(7N_1^2-31N_1N_2+7N_2^2)\) \ .
	\end{align}
	The nonperturbative contribution is \cite{kmt}
	\begin{equation}\label{eq:zmmnp} 
		Z_\text{mm}^\text{np}(N_1,N_2)  = Z_\text{mm}^\text{np,norm}(N_1,N_2) Z_\text{mm}^\text{np,relevant}(N_1,N_2) \ ,
	\end{equation}
	with
	\begin{equation}\label{eq:zmmnp-irrel} 
		Z_\text{mm}^\text{np,norm}(N_1,N_2) = \ri^{N_2} (-1)^{\lfloor\frac{{N_2}}{2}\rfloor}\(\frac{g_s}{m}\)^{(N_1^2+N_2^2)/2} m^{2N_1N_2}\exp\(-\frac{m^3 N_2}{6g_s}\) \ ,
	\end{equation}
	and
	\begin{equation}\label{eq:zmmnp-rel}
		Z_\text{mm}^\text{np,relevant}(N_1,N_2) = (2\pi)^{-(N_1+N_2)/2}G(1+N_1)G(1+N_2) \ .
	\end{equation}		
	Here $G(1+N)$ is the Barnes function. It vanishes when $N = -1,-2,\ldots$, and it has the following asymptotic expansion if $|N|$ is large and $N\not\in \IR_-$
	\begin{equation}\label{eq:Barnes}
		\log (2\pi)^{-N/2}G(1+N) = \zeta'(-1)  + \(\frac{N^2}{2}-\frac{1}{12}\)\log N -\frac{3 N^2}{4} + \sum_{g=2}^\infty \frac{B_{2g}}{2g(2g-2) N^{2g-2}} \ .
	\end{equation}
	In the above expression of $Z_\text{mm}^\text{np}(N_1,N_2)$, the factors in $Z_\text{mm}^\text{np,norm}$ contribute (up to a constant) in the 't Hooft expansion to the ambiguous linear or quadratic terms of the planar free energy. Important are the factors in $Z_\text{mm}^\text{np,relevant}$, which are universal, and which come from volumes of the unitary groups $U(N_1), U(N_2)$.

	Let us turn to the gauge theory side. It has been  proposed  in \cite{Kajiwara:2004ri,betal} that the $H_1$ theory could be related to the Painlev\'{e} II equation. To be precise it was found in \cite{betal} that the $\tau$-solution to the Painlev\'{e} II equation has the form
	\begin{equation}\label{eq:tauII}
		\tau_{II}(T) = S^{-\tfrac{1}{6}+\tfrac{\theta^2}{3}}\sum_{n\in\IZ } \re^{\ri n \rho}\CG(S,\nu+n,\theta) \ ,\quad 8T^3 = 9S^2 \ ,
	\end{equation}
	where  $T$ is the time, $\theta$ a parameter characterising the equation (see equation \eqref{pII}) and $\nu,\rho$ integration constants. The summand $\CG(S,\nu,\theta)$ has the decomposition
	\begin{equation}\label{eq:M-limit-H1}
		\CG(S,\nu,\theta) = {{C}}(S,\nu,\theta) \(1+\sum_{k=1}^\infty \frac{D_k(\nu,\theta)}{S^k}\) \ .
	\end{equation}
	The claim \cite{betal} is then that, with the dictionary
	\be\label{eq:dict-P2G} 
	\ba
		\nu=&-
	 \sqrt{2}\,\ri\, \tilde{a}_D/ \epsilon\\ 
		S=&-{{ 32}\over 3}  (-c)^{3/2}/\epsilon \\
		\theta=& \pm \sqrt{2}\,\ri\, m_{\rm H_1} /\epsilon \ ,
	\ea
	\ee
	the summand $\CG(S,\nu,\theta)$ is identified with the partition function $Z_{\rm H_1}(\tilde{a}_D,m_{\rm H_1},c)$ of the $H_1$ theory in the magnetic frame coupled to the self-dual $\Omega$ background. Note due to the gauge--matrix-model dictionary \eqref{eq:dict-2} $\tilde{a}_D, m_{\rm H_1}$ can identified with the 't Hooft parameters, and $\epsilon$ proportional to $g_s$, therefore we should regard $\nu,\theta$ as counterparts of the ranks $N_1,N_2$ according the dictionary \eqref{eq:dict-P2G}, and \eqref{eq:M-limit-H1} is to be compared with the finite $N$ limit \eqref{eq:M-limit} of the matrix model.
	
	The factor $C(S,\nu,\theta)$ is given in \cite{betal}
	\begin{equation}\label{eq:cdef}
		{ {C}}(S, \nu, \theta)= (2\pi)^{-\nu} G\left(1+\nu + \frac{\theta}{2}\right) G\left(1+\nu - \frac{\theta}{2}\right)\cdot \re^{\ri \nu S + \pi \ri \nu^2/2} S^{-(\nu^2+ \theta^2/4) + 1/6 } 6^{- \nu^2} \ . 
	\end{equation}
	From the point of view of identification with the $H_1$ theory,  	
	the last three terms are irrelevant since they contribute to linear or quadratic terms of the planar free energy and constant term of the genus one free energy, which are ambiguous. %\footnote{The factors $\re^{\pi\ri\nu^2}/2$ and $6^{-\nu^2}$ are in fact a mismatch from the $H_1$ theory, since they correspond to quadratic terms in $F_0$ with non-half-integral constant coefficients, which should be absent.}. 
	 Therefore as in the matrix model we could split $C(S,\nu,\theta)$ by
	\begin{equation}
		C(S,\nu,\theta) = C^{\text{norm}}(S,\nu,\theta) C^{\text{relevant}}(S, \nu, \theta) \ ,
	\end{equation}
	where the essential part reads
	\begin{equation}\label{eq:cdef-rel}
		C^{\text{relevant}}(S, \nu, \theta)= (2\pi)^{-\nu} G\left(1+\nu + \frac{\theta}{2}\right) G\left(1+\nu - \frac{\theta}{2}\right),
	\end{equation}
	while the rest is collected in $C^{\text{norm}}(S,\nu,\theta)$
	\be\label{eq:cdef-irrel}  
		C^{\text{norm}}(S, \nu, \theta)= \re^{\ri \nu S + \pi \ri \nu^2/2} S^{-(\nu^2+ \theta^2/4) + 1/6 } 6^{- \nu^2}\ .
	\ee
	The coefficients $D_k(\nu,\theta)$ can be in principle computed recursively. It is however hard to compute them for higher values of $k$. The first two terms are given in \cite{betal} and they read 
	\be \label{eq:ddef}
	\ba 
		D_1({\nu}, \theta)=&-\frac{\ri}{36} \nu \left(68 \nu^2-9 \theta^2+2\right) \ , \\
		D_2({\nu}, \theta)=&-\frac{289}{162} \nu ^6 +\frac{153 \theta ^2-1159}{324}  \nu ^4 -\frac{81 \theta^4-1584 \theta^2+1084}{2592} \nu ^2 -\frac{\theta ^2 \left(11 \theta ^2-68\right)}{1728}  \ . 
	\ea
	\ee	
	Note that since very few $D_k(\nu,\theta)$ were computed in \cite{betal}, the convergence properties of the large $S$ expansion in \eqref{eq:M-limit-H1} could not be analysed.
	
 One can now easily check that 	
	using the dictionary
	\begin{equation}\label{eq:dict-P2M}
	\begin{aligned}
		S &= -\ri m^3/(6 g_s) \ ,\\
		\nu &= (N_1 - N_2)/2 \ ,\\
		\theta &= -(N_1 +N_2 )\ ,
	\end{aligned}
	\end{equation}
	 the perturbative contributions of the matrix model $g_s^k Z^{(k)}_\text{mm}(N_1,N_2)$ are identified with those of the $H_1$ theory $S^{-k} D_k(\nu, \theta) $, at least for $k=1,2$, while the essential parts of the non-perturbative contributions, namely \eqref{eq:cdef-rel} and \eqref{eq:zmmnp-rel}, agree if one replace $N_2$ with $-N_2$ in \eqref{eq:zmmnp-rel} \footnote{This change of sign in the volume factor is just a minor technicality due to the particular definition of the $\tau$ function of Painlev\'e II in \cite{betal}. If one wishes to match the matrix model without flipping the sign of $N_2$, one can multiply $C(S,\nu+n,\theta)$ in \eqref{eq:tauII},\eqref{eq:M-limit-H1} with 
	 \begin{equation}
	 	{G(1-\nu-\theta/2-n) \over G(1+\nu+\theta/2+n)} = {G(1-\nu-\theta/2)\over G(1+\nu+\theta/2)}\left({\sin \pi (\nu+\theta/2) \over \pi}\right)^n(-1)^{n(n+1)/2} \ ,
	 \end{equation}
	 pulling the $n$-independent ratio of Barnes functions out of the summation, and reabsorbing $\sin(\pi (\nu+\theta/2))/\pi$ into $\rho$ in \eqref{eq:tauII}. We thank Oleg Lisovyy for a discussion on this point.}.
	Hence we will use the notation
	\be\label{eq:myid} 
		 Z_\text{mm}(N_1,N_2) \fallingdotseq \CG(S,\nu,\theta)
	\ee
	where $\fallingdotseq$ means that the equality is only up to the terms that become ambiguous in the 't Hooft expansion (namely \eqref{eq:zmmnp-irrel} and \eqref{eq:cdef-irrel}) and provide we take into account the switch $N_2 \to -N_2$  in \eqref{eq:zmmnp-rel}\footnote{As an additional comment we note that, unlike the matrix models arising in quantisation of mirror curves \cite{bgt,bgt2,mz,kmz,cgum,cgm2}, the one in \eqref{eq:mm2}  is an Hermitian matrix model and it is only perturbatively well defined. Therefore it is unlikely that it can be derived following the geometrical approach  connecting quantum curves and Painlev\'e equations developed in  \cite{Bonelli:2017gdk}. On the other hand this model seems to  fit a bit more naturally into the approach of \cite{Mironov:2017lgl,Mironov:2017sqp} even though in the latter one makes contact with the electric frame and not with the magnetic one which is instead the correct frame  for the model \eqref{eq:mm2}.}.
	\raggedbottom
	Note that the three dictionaries \eqref{eq:dict}, \eqref{eq:dict-P2G}, \eqref{eq:dict-P2M} are consistent if we choose the scaling
	\begin{equation}
		\epsilon = -2\sqrt{2}\,\ri\,  g_s \ .
	\end{equation}
	Furthermore, assuming the dictionary \eqref{eq:dict-P2M} is correct, we can reverse the logic and use the matrix model calculation, which is much easier, to predict $D_k(\nu,\theta)$ for higher $k$. For instance 
		\begin{align}  
		D_3(\nu, \theta) 
		= &\frac{\ri \left(3360-28504 \theta ^2+4270 \theta ^4-99 \theta ^6\right) \nu }{62208} +\frac{\ri \left(899576-700884 \theta ^2+45648 \theta ^4-729 \theta ^6\right) \nu ^3}{279936}\nn
		&+\frac{\ri \left(279464 -47178 \theta ^2 +1377 \theta ^4\right) \nu ^5}{23328} +\frac{\ri \,17 \left(2284 -153 \theta ^2\right) \nu ^7}{5832}+\frac{\ri\, 4913 \nu ^9}{4374}\ .
	\end{align}
	We have computed the expressions of $D_4, D_5, D_6, D_7 $ and $D_8$. We listed some of them in Appendix~\ref{sc:Dk}, while the others are available upon request.
	
	Finally, since many $D_k$ can be computed with relative ease, we can now analyse the convergence property of $\CG(S,\nu,\theta) \propto Z_\text{mm}(N_1,N_2)$. In the cases of $(N_1,N_2) = (1,0), (2,0),(1,1)$, which correspond to $(\nu,\theta) = (1/2,1), (1,2), (0,2)$, $Z_\text{mm}(N_1,N_2)$ can be analytically computed,
	\begin{align}\label{eq:zser}
		Z(1,0) = &{1\over 2 \pi} \sum_{n=0}^{\infty} \frac{2^{3n+\tfrac{1}{2}}}{3^{2n}}\frac{g_s^{n+\tfrac{1}{2}}}{m^{3n+\tfrac{1}{2}}}\frac{\Gamma(3n+\tfrac{1}{2})}{(2n)!}  \ , \nn
		Z(2,0) = & {1\over 2\pi}\frac{g_s^2}{m^2}+{ 1\over (2\pi)^2}\sum_{n=0}^{\infty} \frac{2^{3n+5}}{3^{2n+2}}\frac{g_s^{n+3}}{m^{3n+5}} \cdot\left(\frac{\sqrt{\pi}}{2}\frac{\Gamma(3n+\tfrac{7}{2})}{(2(n+1))!} + \sum_{k=0}^{n}\frac{\Gamma(3k+\tfrac{1}{2})\Gamma(3n-3k+\tfrac{5}{2})}{4(2k+1)!(2n-2k+2)!}\right.\nn
		&\left.\phantom{\frac{\Gamma(3n+\tfrac{7}{2})}{(2(n+1))!}}\times (2k+1)(72k^2-6(17+18n)k+29+66n+36n^2)\right) \ ,\nn
		Z(1,1) = & {\re^{m^2/6}\ri \over 2 \pi^2}\left( 2\pi m g_s + \sum_{n=2}^{\infty}\frac{2^{3n+1}}{3^{2n}}\frac{g_s^{n+1}}{m^{3n-1}}\left(\frac{(-1)^{n+1}15\sqrt{\pi} (n-1)(2n-1)\Gamma(3n-\tfrac{5}{2})}{8\Gamma(2n+1)}\right. \right. \nn
		& \phantom{222222}+\sum_{k=0}^{n-2}\frac{(-1)^k 9}{64(2k)!(2n-2k)!}\Gamma(3k+\tfrac{1}{2})\Gamma(3n-3k-\tfrac{11}{2})(6n-6k-5)(6n-6k-11)\nn
		&\left.\left. \phantom{22222\frac{\Gamma(\tfrac{1}{2})}{\Gamma{2}}}\times(2n-2k-1)(2n-2k-3)(36k^2+(12-36n)k+7-12n)\right)\right) \ .
	\end{align}
	At the level of the $D_k$ coefficients this gives 
		\begin{align}\label{eq:ds}
		D_n(\tfrac{1}{2},-1) &= \frac{(-\ri)^n}{2^{4n}3^{3n}}\frac{(6n)!}{(3n)!(2n)!} \ ,\nn
		D_n(1,-2) &= \frac{(-\ri)^n}{2^{4n}3^{3n}}\cdot\left(\frac{(6n)!}{(3n)!(2n)!}+2\sum_{k=0}^{n-1}\frac{(6k)!}{(3k)!(2k)!}\frac{(6n-6k-2)!}{(3n-3k-1)!(2n-2k)!}\right.\nn
		&\left.\phantom{\frac{(6n-6k-2)!}{(3n-3k-1)!(2n-2k)!}}\times(72k^2-6(18n-1)k+36n^2-6n-1)\right)\nn
		D_n(0,-2) &=\frac{(-\ri)^n}{2^{4n}3^{3n}}\cdot\left((-1)^{n+1}120\frac{(6n-6)!}{(2n)!(3n-3)!}(n-1)(2n-1) \right. \nn
		&+ 576\sum_{k=0}^{n-2}(-1)^k\frac{(6k)!}{(3k)!(2k)!}\frac{(6n-6k-1)!}{(3n-3k-6)!(2n-2k)!}(6n-6k-5) \nn
		&\left.\phantom{\frac{(6k)!}{(3k)!(2k)!}}\times(2n-2k-1)(2n-2k-3)(36k^2-(36n-12)k-12n+7)\right) \ .
	\end{align}
	All three series in \eqref{eq:zser} are divergent. In fact the coefficients of $g_s^n/m^{3n}$, denoted by $a_n$, in all three series have the asymptotic behavior
	\begin{equation}
		|a_n| \sim 6^n n! \ , \quad n\rightarrow \infty\ .
	\end{equation}
	We conjecture this to be always the case for any values of $N_1,N_2$.  As a result, $D_n(\nu,\theta)$ for any value of $\nu,\theta$ has the asymptotic behavior
	\begin{equation}\label{eq:Dk-asymp}
		D_n(\nu,\theta) \sim (-\ri)^n\, n! \ ,\quad n\rightarrow \infty \ .
	\end{equation}
	Hence, unlike the series expansions appearing in the short time solution of Painlev\'e equations \cite{gil,ilt,gil1}, those at long time \cite{betal,Its:2016jkt} seems to suffer from divergence problems. This is somehow expected since generically also  the classical special functions  have  divergent long-distance expansion\footnote{We would like to thank Oleg Lisovyy for a discussion on this point.}.
	We will see in the next section that the sum over all integer shifts in the $\tau$ function \eqref{eq:tauII}  can be interpreted as summing over all instanton sectors and that  the overall normalisation factor \eqref{eq:cdef-irrel}  leads to  the correct instanton action as extracted from the analysis of the large order behaviour namely \eqref{eq:Dk-asymp}.

	\section{Trans-series solution to Painlev\'e II equation} \label{sec:pa}

	Eq.~\eqref{eq:M-limit-H1} is given in \cite{betal} as the $\tau$-function solution to the Painlev\'{e} II equation. The Painlev\'{e} II equation reads
	\be \label{pII}
		q''=2 q^3+T q+{1\over 2}-\theta \ ,
	\ee
	where the derivatives of $q$ are w.r.t.~$T$ while $\theta$ is a  parameter. This is a second order differential equation and its solution depends on two integration constants which, by following the notation of \cite{Its:2016jkt,betal},  we denote by 
	\be (\nu, \rho)\ . \ee
	The corresponding $\tau$ function is defined by \cite{betal}
	\be \label{tauii}
		{\rd \over \rd T} \log \tau= \frac{1}{2} q'^2+\theta  q-\frac{1}{2} q^4-\frac{1}{2} T q^2-\frac{T^2}{8}\ . 
	\ee
	In \cite{Its:2016jkt,betal} it was found that the $\tau$ function associated to a generic solution to the Painlev\'{e} II equation in the large $T$ limit along the rays $ {\rm arg}(T)=0, \pm {2\pi\over 3}$	can be written as
	\be\label{eq:tauLy} 
		\tau(T,\nu, \rho, \theta)=\sum_{n \in \IZ} \re^{\ri n \rho} \CG(S,\nu+n, \theta) \ ,
	\ee
	where $\CG(S,\nu, \theta)$ is given in \eqref{eq:M-limit-H1} and we use the change of variables
	\be 
		8 T^3=9 S^2 \ .
	\ee

	In this section we argue that \eqref{eq:tauLy} can be reproduced by a trans-series solution to the Painlev\'{e} II equation. We will check this explicitly along the slice\footnote{Note that in the matrix model perspective this slice corresponds to the one-cut phase.}
	\begin{equation}\label{eq:slice}
		\nu = -\theta/2 \ ,
	\end{equation}
	where the trans-series solution can be easily constructed by following \cite{mmnp}.
	Let us make the following trans-series ansatz for the function $q(T)$\footnote{This ansatz is obtained from \cite{mmnp} where the case $\theta=1/2$ was studied.}	
	\begin{equation}\label{ansatz}
		q(T, \sigma, \theta)=\sum_{n \geq 0}\sigma^n \re^{-n A/x} q^{(n)}(T)=x^{-1/3}\sum_{n\geq 0} \sigma^n \re^{-n A/x}x^{n \beta}\sum_{g\geq 0} u_g^{n}x^g \ ,
	\end{equation}
	where
	\begin{equation}
		x=T^{-3/2} , \quad \beta=1-{\theta}, \quad A=-  {2 \sqrt{2}\over 3} \,\ri \ .
	\end{equation}
	Here $q^{(0)}(T)$ is the perturbative sector, $q^{(n\geq 1)}(T)$ different instanton sectors, and $A$ is interpreted as the instanton action.
	  By plugging this ansatz in \eqref{pII} we can compute all the coefficients. We find for instance for the perturbative part
	\be 
		u_0^0=-\frac{\ri}{\sqrt{2}}, \quad u_1^0=\frac{1}{4}-\frac{\theta }{2}, \quad u_2^0=-\frac{\ri (12 (\theta -1) \theta +5)}{16 \sqrt{2}}, \quad \cdots 
	\ee
	and the first instanton sector
	\be \ba  
		u_1^1=& -\frac{\ri (3 \theta  (7 \theta -10)+14)}{12 \sqrt{2}} u_0^1\ , \\
		u_2^1= &\frac{1}{576} (3 \theta  (\theta  (3 (222-49 \theta ) \theta -1045)+844)-799)u_0^1\ , \\
	\cdots 
	\ea  \ee
	where $u_0^1$ is a free parameter.  Likewise for the second instanton sector we have 
	\be \ba 
	u_0^2=&\frac{ \ri }{\sqrt{2}}(u_0^1)^2\\
	u_1^2= &\frac{1}{12} (21 (\theta -2) \theta +26)(u_0^1)^2\\
	\cdots
	\ea \ee
	It is then easy to check that the trans-series solution \eqref{ansatz} when plugged into the definition of $\tau$ function \eqref{tauii} reproduce the solution \eqref{eq:tauLy},\eqref{eq:M-limit-H1},\eqref{eq:cdef},\eqref{eq:ddef} in the slice \eqref{eq:slice}, as long as we choose
	\begin{equation}
	\begin{gathered}
		\sigma = \re^{\ri n \rho} \ ,\\
		u_0^1= -\frac{2^{\frac{5 \theta }{2}-3} e^{-\frac{1}{2} i \pi  \theta } \Gamma (1-\theta )}{\pi } \ . 
	\end{gathered}
	\end{equation}
	In this identification the summation appearing in \eqref{eq:tauLy} coincides with the sum over different instanton sectors in \eqref{ansatz}. 
	In particular the weight $\re^{\ri (\nu+n) S}$ in the sum  \eqref{eq:tauLy}  is in fact related to the instanton action $\re^{-nA/x}$ in the \eqref{ansatz}, from which we read off the instanton action $-\ri$, consistent with the asymptotic behavior of $D_n(\nu,\theta)$ when $n\rightarrow \infty$ given in \eqref{eq:Dk-asymp}. In the matrix model language this corresponds to the factor $ \exp\(m^3 (-N_2+n)/(6g_s)\) $ in \eqref{eq:mm2}.Furthermore this shows explicitly  how the $n^{\rm th}$ instanton sector in \eqref{ansatz} is completely determined by the same function as the perturbative sector, namely the term $n=0$ in \eqref{eq:tauLy} or \eqref{ansatz}.

	{ It is interesting to compare more in details the above solution, which in turns is a rewriting of \cite{betal,Its:2016jkt} in the trans-series language, with the solution at $\theta=1/2$ proposed in \cite{mmnp}  \footnote{ The parameter $\kappa$ in \cite{mmnp} is related to our parameter $T$ as $T=-2^{1/3}\kappa$}. There is a subtle difference between the two which is due to a different choice of the perturbative sector. Indeed if we plug an ansatz of type \eqref{ansatz}  for a generic $A$ in \eqref{pII}  we obtain the following equations for $u_0^0$ and $A$:
	\be \ba  -u_0^0 - 2 (u_0^0)^3=0,  \\
	9 A^2-24 ( u_0^0)^2-4=0.\ea\ee
	Since we want to make contact with the solution of \cite{betal,Its:2016jkt} we chose
	\be \label{our}u_0^0=-\frac{\ri}{\sqrt{2}}.\ee
On the contrary in \cite{mmnp} the perturbative sector was chosen to be  
\be\label{marcos} u_0^0=0.\ee
This implies $A = \pm 2/3 $  in which case the factor $\re^{-{A} T^{3/2}}$
is suppressive. For the choice \eqref{our} one has insead
${A} = \pm 2\sqrt{2}/3 \ri$, which makes contact with the solution of \cite{betal,Its:2016jkt} we previously discussed, in
which case the factor $\re^{-{A} T^{3/2}}$ is oscillatory.
	  }

		%The generic solution \eqref{eq:tauLy} with two parameters away from the slice \eqref{eq:slice}, however, to our knowledge  does not an existing counterpart in the trans-series literature. 
	%
Using the identification \eqref{eq:myid} with the cubic matrix model, we can write 
	\begin{equation}\label{eq:taumm}
		\tau(T,\nu,\rho,\theta) \fallingdotseq \sum_{n\in \IZ} \re^{\ri n \rho} Z_\text{mm}(N_1+n,N_2-n) \ .
	\end{equation}
	As explained for instance in \cite{msw},  
	\be 
		Z_\text{mm}(N_1+n,N_2-n)
	\ee 
	can be interpreted in a multi-cut matrix model in terms of eigenvalue tunneling between different branch cuts. Hence the sum in the $\tau$ function of Painlev\'e equation is interpreted as a sum over all possible tunneling of eigenvalues. From that perspective the sum over integers appearing in \eqref{eq:tauII} is similar to the sum over filling fractions appearing in the matrix model literature \cite{bde,em,eynard}.

 If we think of the form of $\tau$-function \eqref{eq:tauII} as a trans-series summing over all instanton sectors, it makes sense to look at the Borel resummation of the perturbative sector with $n=0$ in \eqref{eq:tauII} as a possible means to reproduce exact solutions to the PII equation. In particular 
	we notice that in the three calculated examples \eqref{eq:ds} with $(\nu,\theta) = (1/2,-1),(1,-2),(0,-2)$, the asymptotic series in \eqref{eq:M-limit-H1} are Borel summable. We expect this to be the case also for generic values of $(\nu,\theta)$. Let us consider the case with $(\nu,\theta) = (1/2,-1)$, the Borel resummation
	 \be 
		 { \mathcal{B} }\left(S\right) =\int_0^{\infty} \rd z \re^{-z}\sum_{n\geq 0}{D_n (1/2,-1)\over n!}{z^n\over S ^n}
	\ee
	 can be performed exactly and it yields
	\begin{equation}
		{ \mathcal{B} \left(S\right)} = \re^{-\tfrac{\ri S}{2}} \sqrt{\frac{S}{\ri\pi}}\,K_{1/3}\(-\frac{\ri S}{2}\) \ ,
	\end{equation}
	where $K_{1/3}(\bullet)$ is the modified Bessel function of the second kind with order $1/3$. Hence the corresponding $\tau$ function is
	\be 
		\tau (T,1/2, -1 )=\frac{\re^{\frac{\ri S}{2}+\frac{\ri \pi }{8}} \mathcal{B} \left(S\right)}{2^{3/4} \sqrt[4]{3} \sqrt{\pi } \sqrt[6]{S}}
		=  6^{-1/12}\, \re^{\frac{\ri \pi }{24}} \text{Ai}\left(2^{-1/3} \re^{-\frac{ \ri \pi }{3}} T \right), \qquad  8 T^3=9 S^2 \ ,
	\ee
	where $\text{Ai}(\bullet)$ is the Airy function.
	It is easy to verify that 
	\be 
		{\rd \over \rd T} \log \tau (T,1/2, -1 )
	\ee
	satisfies the PII equation in the $\sigma$ form  { for $\theta=-1$ and with asymptotic expansion characterized by $\nu=1/2$}. We have hence recovered the well known Airy solution to the PII equation. 
	By combining Borel resummation with the matrix model representation of the coefficients $D_k(\nu,\theta)$, one can in principle construct more complicated closed form solutions for other values of $(\nu,\theta)$.  Note also that the Airy function can be simply obtained by changing the integral contour in the one-cut phase of the cubic matrix model in line with the non-perturbative matrix model formulation of \cite{bde,eynard}. For the two-cut case it would be interesting to further compare the Borel resummation of $D_k(\nu,\theta)$ for generic values of $(\nu, \theta)$ with the approach of \cite{Felder}. This is left for further investigations.

	\section{Argyres Douglas theories and quantum mechanics}\label{sec:qm}
 In this section  we focus on the Argyres-Douglas theories in the Nekrasov-Shatashvili limit \cite{ns} where the two regulators are given by
	\be 
		\epsilon_1=\hbar, \quad \epsilon_2\to 0 \ .
	\ee
	This limit is closely connected to the self-dual limit studied in the previous sections through the 
	 blowup equations \cite{ggu}. Interestingly both these limits  admit an operator theory interpretation which,  in case of  pure $SU(N)$ theory, was worked out in \cite{ns,mirmor2} for the NS background and in \cite{bgt,bgt2} for the self-dual background.
	 
	In this section we  discuss two types of AD theories: the $H_0$ and the $H_1$ theories.  
	They correspond to special limits in the moduli spaces of the four dimensional $\CN=2$ $SU(2)$ SQCD with $N_f=1$ and $N_f=2$ where mutually nonlocal dyons become massless simultaneously \cite{ad2, ad1}. Let us still focus on the magnetic frame. Then the free energy $ \CF$ of these theories in the NS limit display the following perturbative behaviour 
	\be 
		\CF^{\rm D}(a_{\rm D})=\sum_{g\geq 0} \epsilon^{2g-2}  \CF_g^{\rm D}(a_{\rm D})\ ,
	\ee
	where the NS free energies $\CF_g^{\rm D}(a_D)$ can be computed recursively by using the NS limit of the refined holomorphic anomaly equation  \cite{hk,kwal} \footnote{In the notation of section \ref{sec:mmandad} these correspond to $F_{0,n}$.}. 
	Note that the planar free energies in the NS and the self-dual limit are the same, and we will simply denote it by $ F_0^{\rm D}$ instead of  $ \CF_0^{\rm D}$.
	
	We find that the NS limit of the $H_0$ and the $H_1$ theories capture the spectral properties of certain QM models. To be specific, the $H_0$ theory corresponds to the QM model with the cubic potential, while the $H_1$ theory the QM model with the double well potential\footnote{A connection between some  aspects of these two QM models and some invariants of the $\CN=2$ $SU(2)$ SQCD with $N_f=1$ and $N_f=2$ respectively was also discussed in \cite{Basar:2017hpr}.}. In fact they are just two examples of a larger story which relate the quantization of four dimensional $SU(2)$ Seiberg-Witten spectral curves \cite{ns,mirmor}  \footnote{ Here we are only concerned with 4d theories with gauge group $SU(2)$. The quantization of Seiberg-Witten curves for 5d $\CN=1$ gauge theories were discussed in \cite{acdkv, Nekrasov:1996cz} with the WKB approximation, while the exact answers were proposed in \cite{ghm,cgm2}, see also \cite{gkmr,Wang:2015wdy,ggu,huang1606,Grassi:2017qee}.
	} to the all order WKB solutions \cite{Dunham1932:WKB} (see \cite{Bender1977:WKB,Galindo1990:WKB} for a clear presentation) of QM models with polynomial potentials. Let us take a QM model with Hamiltonian
	\be 
		H=-\partial_x^2+V(x) \ ,
	\ee 
	where $V(x)$ is a polynomial in $x$. Define the spectral curve
	\begin{equation}
		\CC_\text{QM}: \quad p^2 + V(x) = E \ .
	\end{equation}
	The perturbative energy levels of the QM system are solved from the quantum period of the cycle on  $\CC_\text{QM}$ associated to the classically accessible region of the potential $V(x)$, while the nonperturbative corrections are encoded in the Voros multiplier \cite{voross,Voros1981,Silverstone1985:WKB}, the quantum period of the cycle on $\CC$ associated to the classically forbidden region. It was conjectured and verified in \cite{csm} (see also \cite{cms,cesv1}) through examples of cubic potential and double well potential that the Voros multiplier together with the perturbative energy levels define the quantum free energy as an analogue of NS free energy and it can be solved from the NS holomorphic anomaly equations\footnote{This algorithm can be justified to some extent by \cite{ag}, which demonstrates that under the assumption that the quantum periods under a symplectic transformation behave like classical periods, the quantum free energy satisfies the NS holomorphic anomaly equations.}. This gives a relatively easy and systematic way to compute the Voros multiplier of a QM model. From this perspective, if we can find a 4d $\CN=2$ theory whose SW curve coincides with the spectral curve $\CC_\text{QM}$ of the QM model, and whose SW differential is
	\begin{equation}\label{eq:differential}
		\Omega_\text{SW} = y(x) \rd x \ ,
	\end{equation}
which coincides with the exponential of the WKB solution to the QM model in the leading order, then naturally the periods of the QM model can be identified with those of the 4d theory, and the quantum free energy with the NS free energy. As advertised before, we will {illustrate} this idea for the QM models with the cubic potential and the double well potential.  This identification of QM models with $\CN=2$ theories gives a gauge theory justification for the algorithm proposed in \cite{csm}.  Note also that a connection between QM models with monic potentials and AD theories was noted in \cite{ito-shu} in the context of ODE/IM correspondence \cite{Dorey:2007zx,Dorey:1998pt}.

	\subsection{The NS limit of $H_0$ theory and the cubic oscillator}\label{sec:cubic}
	
	\begin{figure}
		\centering
		\includegraphics[width=0.5\linewidth]{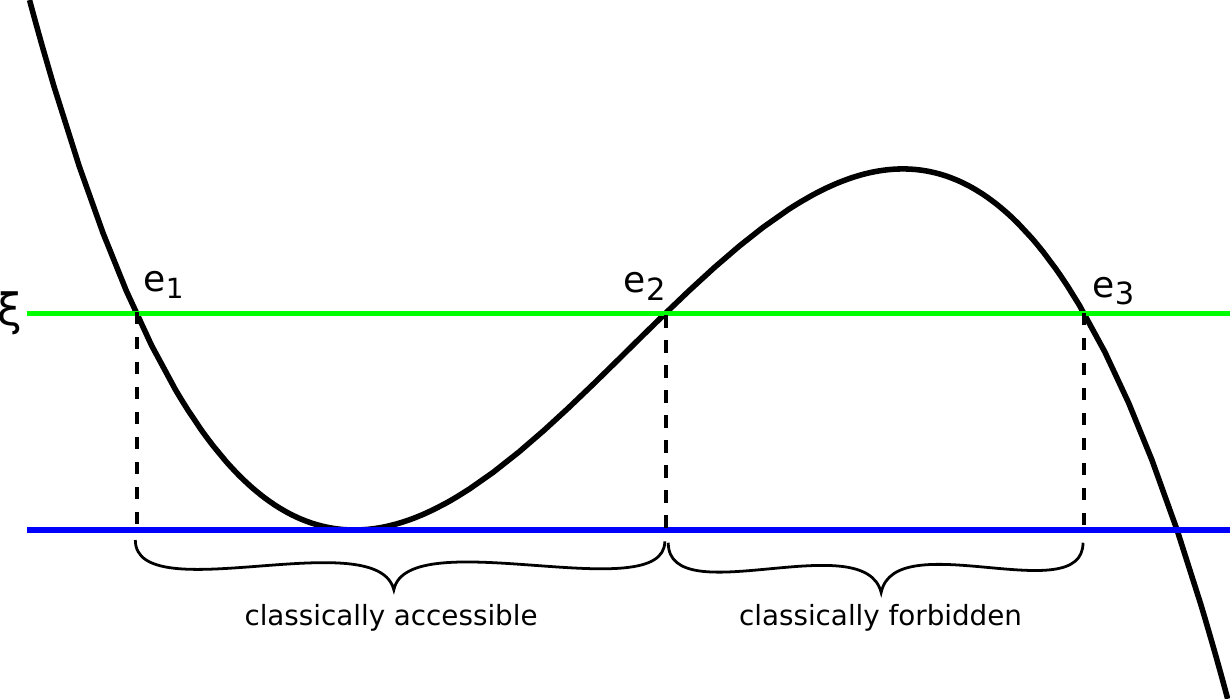}
		\caption{The classically accessible and forbidden regions of a cubic potential.}\label{fg:cubic}
	\end{figure}
		
	The  cubic oscillator is a one-dimensional quantum mechanical system characterised by the following potential
	\be \label{eq:Vcubic} 
		V(x)={x^2\over 2}- g x^3 \ .   
	\ee
	The spectral curve $\CC_\text{QM}$ used to compute quantum periods and quantum free energy is \cite{csm}
	\be 
		\CC_\text{QM}: \quad y^2=2\xi-x^2+2gx^3 \ ,
	\ee
	where $\xi$ is identified with the energy of the QM model.
	We perform the following linear change of variables
	\be 
		x \to (2g)^{-1/3}x+1/(6g)  \ .  
	\ee
	 then $\CC_\text{QM}$ becomes
	\be \label{eq:sp-H0}
		y^2=x^3  - 3c x + u \ ,
	\ee
	where
	\be \label{dict}  
		u=2 \xi -\frac{1}{54 g^2} \ , \quad c ={\frac{1 }{3^2 (2g)^{4/3}} } \ ,
	\ee
	which is precisely the SW curve for the $H_0$ theory (see eq (4.6) of \cite{ad1}, or (4.10) of \cite{betal}). $u$ is the Coulomb modulus, while $c$ is the scale parameter that controls the deformation away from the conformal point. In addition the SW differential of the $H_0$ theory is indeed of the form 
	\eqref{eq:differential}.
	
	The QM model is studied in the semi-classical limit $\hbar \rightarrow 0$, in which case the period of the SW differential integrated around the classically accessible region (see Fig.~\ref{fg:cubic}) should shrink to zero \cite{csm}. In the AD theory, this corresponds to a conifold point of the moduli space. The discriminant of the SW curve \eqref{eq:sp-H0} is
	\begin{equation}
		\Delta = 27 (u^2-4c^3) \ ,
	\end{equation}
	and thus two conifold points exist
	\begin{equation}
		u_\pm = \pm 2 c^{3/2} \ .
	\end{equation}
	They correspond to the vanishing of the cycles around either the classically accessible region or the forbidden region. Since these two regions are exchanged to each other by $x\rightarrow -x$, the two conifold points are on equal footing, and we can choose either one. In the magnetic frame around the conifold point, say, $u_+$, the period $a_D$ is the good local coordinate and it has in general the form (up to a normalization constant)
	\begin{equation}
		\frac{\rd a_D}{\rd u} {~ \propto ~} \frac{1}{2\pi}\int_{e_1}^{e_2}\frac{\rd x}{y} \ ,
	\end{equation}
	where we have assumed $e_1, e_2, e_3$ to be the three branch points of \eqref{eq:sp-H0} from left to right, as shown in Fig.~\ref{fg:cubic}. The integral above can be written as a hypergeometric function
	\begin{equation}
		\frac{\rd a_D}{\rd u} { ~ \propto ~} \frac{1}{2}(e_3 - e_1)^{-1/2} {}_2F_1\(\frac{1}{2},\frac{1}{2};1;z\) \ ,
	\end{equation}
	with
	\begin{equation}
		z = \frac{e_2 - e_1}{e_3 - e_1} \ .
	\end{equation}
	Then following the same technique in \cite{Masuda:1996xj}, we can transform the hypergeometric function to a form symmetric in $e_{1,2,3}$. The result is 
	\begin{equation}
		\frac{\rd a_{\rm D}}{\rd u} { ~ \propto ~} \frac{1}{2}(-\cD)^{-1/4} {}_2F_1\(\frac{1}{12},\frac{5}{12};1;-\frac{27\Delta}{4\cD^3}\) \ ,
	\end{equation}
	where $\Delta$ is the discriminant, while
	\begin{equation}\label{eq:cD-1}
		\cD = -{1 \over 2} \sum_{i<j}(e_i-e_j)^2 \ ,
	\end{equation}
	which is $-9c$ in the case of \eqref{eq:sp-H0}. Therefore, we find
	\be\label{eq:daDdu}
		{\rd a_{\rm D}  \over \rd u}=\frac{\, _2F_1\left(\frac{1}{12},\frac{5}{12};1;-\frac{u^2}{4 c^3}+1\right)}{{2^{1/3}} \cdot 3^{1/2} c^{1/4}} \ .
	\ee
	Likewise the prepotential $F_0^{\rm D}(a_D)$ is computed in \cite{Masuda:1996xj, betal} and it reads  (up to a normalization constant)
	\be  \ba 
		F_0^{\rm D}=&{a_D ^2\over 4}\left(\log{a_D \over 48\ 2^{2/3} 3^{3/2} c^{5/4}}-{3}\right) +\frac{4}{5}  2^{2/3}3^{3/2} c^{5/4}a_D+\frac{47 a_D^3}{48\ 2^{2/3} 3^{3/2} c^{5/4}}\\
		&+\frac{7717 a_D^4}{248832 \cdot 2^{1/3} c^{5/2}}+ \mathcal{O}(a_D^5) \ .
	\ea\ee
	By using the dictionary \eqref{dict} with $g=1$ we find (after normalization)
	\be \label{eq:adsmall}
		a_{\rm D}=\xi+\frac{15 \xi ^2}{4}+\frac{1155 \xi ^3}{16}+\frac{255255 \xi ^4}{128}+\cdots\ ,
	\ee	
	and 
	\be  
		F_0^{\rm D}={a_D ^2\over 2}\left(\log{a_D \over 8}-{3\over 2}\right) +\frac{2 a_D}{15}+\frac{47 a_D^3}{8}+\frac{7717 a_D^4}{128}+ \mathcal{O}(a_D^5) \ .
	\ee
	They are precisely the period associated to the classically accessible region and the planar component of the quantum free energy of the cubic oscillator given in \cite{csm} (eq.~(3.22) and eq.~(3.23) respectively). Next the genus one quantum free energy is computed by \cite{csm} 
	\begin{equation}\label{eq:NSF1}
		\CF^{\rm D}_1(a_D) = -\frac{1}{24}\log \Delta \ ,
	\end{equation}
	which is exactly how one would compute the genus one NS free energy for the gauge theory. In fact, both the quantum free energies of the cubic oscillator and the NS free energies of the $H_0$ theory are computed by the NS holomorphic anomaly equations, and they share the same initial conditions. Therefore the all order WKB solutions to the cubic oscillator are captured by the $H_0$ theory coupled to the $\Omega$ background in the NS limit.

Finally, we comment on the symmetry between the classically accessible region and forbidden region, which leads to that the quantum free energies are invariant under an $S$-transformation (see \cite{csm} for more details). When translated to the gauge theory, it means the two conifold points $u_\pm$ are completely dual to each other, and the NS free energies expanded around these two singular points are identical.

	\subsection{The NS limit of  $H_1$ theory and the double well potential}
	
	\raggedbottom
	
	\begin{figure}
		\centering
		\includegraphics[width=0.5\linewidth]{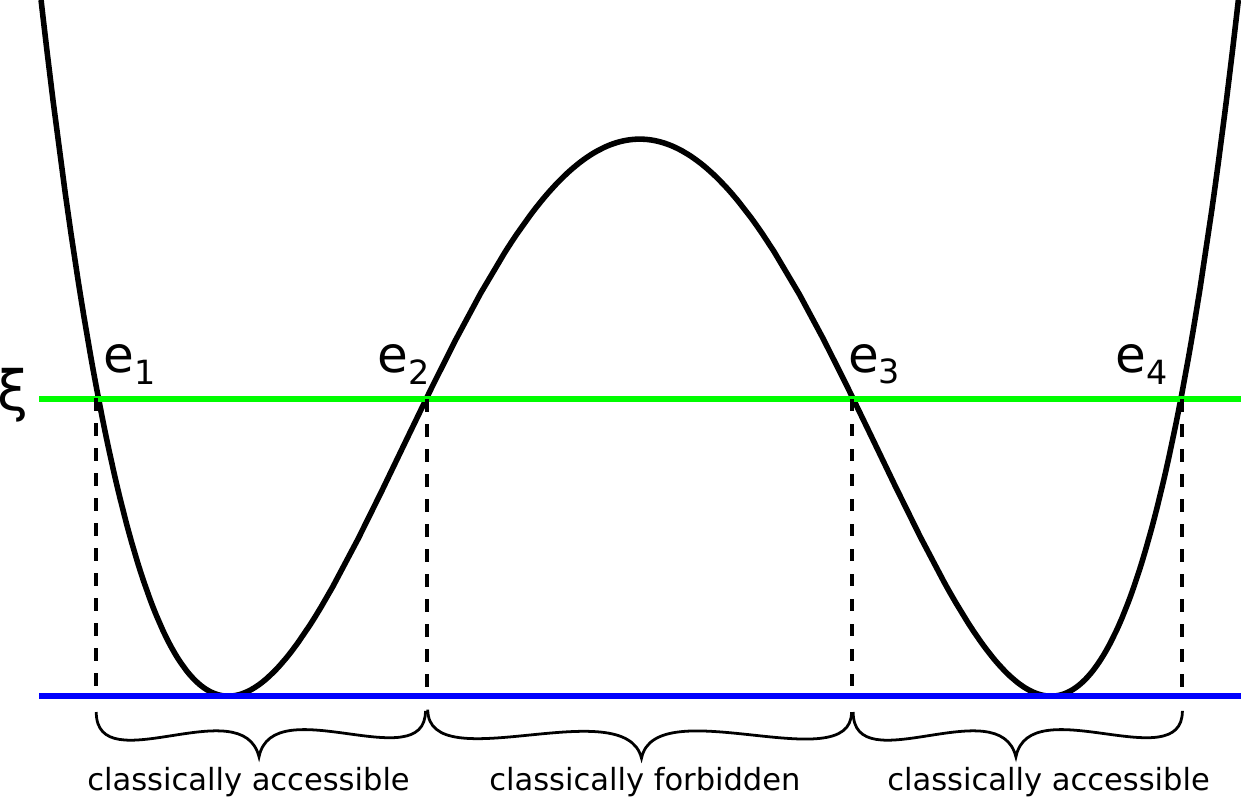}
		\caption{The classically accessible and forbidden regions of a double-well potential.}\label{fg:double-well}
	\end{figure}	
	
	Let us consider the QM model with the double well potential
	\be 
		V(x)={x^2\over 2}(1+g x)^2 \ .
	\ee
	The associated spectral curve is, after scaling and shifting to put it in a symmetric form \cite{csm}
	\begin{equation}\label{curv}
		\CC_\text{QM}:\quad y^2 = (x^2-a_+^2)(a_-^2 - x^2) \ ,
	\end{equation}
	where
	\be 
		a_{\pm}^2=\frac{1}{4} \left(\pm \sqrt{32\xi}  + g^{-1} \right)\ .
	\ee

	Through the scaling
	\be 
		x \to \frac{\re^{\frac{\ri \pi }{4}} x}{\sqrt{g}}  \ ,
	\ee
	we can write \eqref{curv} as
	\begin{equation}\label{ym} 
		y^2 = x^4 + 4c x^2 + 2m x + u \ ,
	\end{equation}
	where 
	\be \label{dic2}
		u=\frac{1}{16} \left(32 \xi -g^{-2}\right) \ ,\quad
		c=\frac{\ri}{8} \,g^{-1}, \quad  m=0  \ ,
	\ee
	and this is precisely the  SW curve \eqref{eq:ch1} for the $H_1$ theory, and as we have seen in Sec.~\ref{sec:H1} the SW differential $\Omega_{H_1}$ of this theory is of the form \eqref{eq:differential}.
	
	In the QM model, the classically accessible regions are indicated in Fig.~\ref{fg:double-well}. In the semi-classical limit, the period of $\Omega_{H_1}$ around this region vanishes, corresponding to a conifold point of the moduli space of the SW curve. We have seen in Sec.~\ref{sec:H1} that there are three conifold points $u=u^{(i)}, i=1,2,3$. Among them, $u^{(2,3)}$ are due to the vanishing of either of the two branch cuts, and they correspond to the two classically accessible regions of the double-well potential, while $u^{(1)}$ which is due to the merger of the two branch cuts, corresponds to the classically forbidden region. Therefore, the semi-classical limit corresponds to either $u^{(2)}$ or $u^{(3)}$.
	
	In the magnetic frame in the vicinity of the singular point $u^{(2)}$, the period $a_D$ is the locally good coordinate, and it can be computed by using the generic formula in \cite{Masuda:1996xj} (up to a normalization constant)
	\begin{equation}
		\frac{\rd a_D}{\rd u} = \frac{1}{2}(-\cD)^{-1/4} {}_2 F_1\(\frac{1}{12}, \frac{5}{12};1; -\frac{27 \Delta}{4 \cD^3}\) \ ,
	\end{equation}
	where $\Delta$ is the discriminant, and $\cD$ is given by\footnote{Note this is different from the expression of $\cD$ for the cubic form. Besides, the expression for $\cD$ in terms of roots in \cite{Masuda:1996xj} is incorrect, while the expression in terms of equation coefficients is correct.}
	\begin{equation}
		\cD = -\frac{1}{2}\((e_1 - e_2)^2(e_3 - e_4)^2+(e_1 - e_3)^2(e_2 - e_4)^2+(e_1 - e_4)^2(e_2 - e_3)^2\) \ ,
	\end{equation}
	with $e_i, i=1,2,3,4$ being the four branch points (see Fig.~\ref{fg:double-well}). In the case of \eqref{ym}, we have
	\be 
		\frac{\rd a_D}{\rd u}= \frac{\, _2F_1\left(\frac{1}{12},\frac{5}{12};1;\frac{27 u \left(4c^2- u\right)^2}{\left(4c^2+3 u\right)^3}\right)}{2\sqrt{2} (3 u+4 c^2)^{1/4}}\ .
	\ee
	After applying the dictionary \eqref{dic2} (g is set to 1) and expanding w.r.t.\ $\xi$, we get (after normalization)
	\begin{equation}
		a_D = \xi +3 \xi ^2+35 \xi ^3+\frac{1155 \xi ^4}{2}+\frac{45045 \xi ^5}{4}+\frac{969969 \xi ^6}{4} + \CO(\xi^7)\ .
	\end{equation}	
	The prepotential $F_0^{\rm D}(a_D)$ can be found in \cite{Masuda:1996xj, betal}, and when $m=0$ it reads
	\be \label{f02} 
		F_0^D= a_D^2 \log \left(\frac{a_D}{2}\right)+\frac{a_D}{3}-\frac{3 a_D^2}{2}+\frac{17 a_D^3}{3}+ \frac{125 a_D^4}{4}+\mathcal{O}(a_D^5) \ .
	\ee
	They indeed agree with the period and genus zero quantum free energy given in \cite{csm}.
	Furthermore, the genus one quantum free energy is computed by \eqref{eq:NSF1} \cite{csm}, and this is how one would compute the genus one NS free energy for the gauge theory. Higher genus free energies would also agree, since they are computed both in the QM model with double well potential and in the $H_1$ theory from the NS holomorphic anomaly equations and they share the same initial data. As a result, the all order WKB solutions to the double well QM model are captured by the $H_1$ theory coupled to the $\Omega$ background in the NS limit.
	
	\section{Summary and open questions} 
		
	It is an interesting problem to look for matrix model representations of supersymmetric gauge theories. It has been proposed in \cite{tc12,Rim:2012tf} that many Argyres-Douglas (AD) theories can be represented by hermitian matrix models with rational/logarithmic potentials. In this paper we show in detail that the well-studied $\beta$-deformed cubic matrix model in the generic two-cut phase computes the partition function of the $H_1$ AD theory coupled to the $\Omega$ background in the magnetic frame. Then we further extend this relation to integrable non-linear ODEs. According to the Painlev\'e/gauge correspondence \cite{betal}, the $H_1$ AD theory in the self-dual $\Omega$ background is expected to compute the $\tau$ function of the Painlev\'e II equation. By combining these two observations we showed that, in the non-deformed $\beta=1$ case, the two-cut cubic matrix model computes the $\tau$ function of the Painlev\'e II equation. Using this connection with matrix model, we studied in detail the $\tau$ function solution to Painlev\'e II proposed in \cite{betal,Its:2016jkt}, and we found that the summand \eqref{eq:M-limit-H1} appearing in this solution is in fact an asymptotic series with zero radius of convergence. Relatedly, when considering the $\tau$ function one has to sum over all integer shifts as illustrated in equation \eqref{eq:tauII}. We found that, from the resurgence perspective, this summation corresponds to a sum over trans-series which, in the matrix model language,  amounts to a sum over all possible eigenvalues tunnellings or a sum over all filling fractions similar to \cite{bde,em,eynard}. Furthermore our analysis also shows explicitly how the $n^{\rm th}$ instanton sector of the PII solution in \eqref{ansatz} is completely determined by the same function as the perturbative sector, namely the term $n = 0$ in \eqref{eq:tauLy} which is computed by the two-cut cubic model via \eqref{eq:myid} and \eqref{eq:dict-P2M}.
	% \eqref{eq:tauLy} or \eqref{ansatz}.
	
	We note that the connection between Painlev\'e equations and
        hermitian matrix models has been explored before in the
        literature in particular in connection with 2d gravity
        \cite{Douglas:1989ve,Brezin:1990rb,Gross:1989vs,Kharchev:1991cy}. See
        \cite{DiFrancesco:1993cyw} for a more exhaustive list of
        references. For instance in \cite{dss,cmo} it was found that
        the quartic matrix model in the double scaling limit is
        related to the Painlev\'e II equation \eqref{pII} at
        $\theta=1/2$.  Likewise the Gross--Witten--Wadia model also
        makes contact with Painlev\'e II in a particular double
        scaling limit. See for instance \cite{mmnp} and reference
        therein. Other models with external fields were also
        introduced in this context; see for instance
        \cite{Mironov:1994mv} and reference therein. From that
        viewpoint what distinguishes our matrix model representation
        for Painlev\'e II from the previous ones is that, after
        summing over all possible eigenvalues tunneling, it computes
        the $\tau$ function of the Painlev\'e II without taking the
        double scaling limit nor adding any external fields and that
        it is valid for generic values of $\theta$ and integration
        constants $(\nu, \rho)$.
	
	Finally we explored the Nekrasov-Shatashvili phase of the $H_0$ and $H_1$ AD theories and we showed that they determine the spectral properties of corresponding quantum mechanical systems with cubic and double well potentials respectively. This provides a gauge theory justification for the all order WKB solutions from holomorphic anomaly equations proposed in \cite{csm} for these two quantum mechanical models. 
	
	There are still many open questions that remain to be addressed. One of the obvious questions is whether one can find a similar matrix model representation for the solutions to PI and PIV presented in \cite{betal,Lisovyy:2016qig}. Besides, the matrix model for the $H_1$ theory is studied in the weak coupling limit, very far away from the conformal point. It would be interesting to explore the strong coupling limit $g_s/m\to 0$, probably following \cite{Cordova:2016jlu}, and see if one can construct in this way  solutions to  Painlev\'e II for small times. Furthermore it would be interesting to study in more detail the Borel resummation of the perturbative sector of the $\tau$-function, whose coefficients can be computed from the two-cut cubic matrix model, similar to what we have done at the end of section \ref{sec:pa}.  
	%This would also allow us to analyse closer the non--perturbative effects in $H_1$ theory
	%	 Another related question is  to further investigate the Borel resummation of the $D_k$ at generic $(\nu, \theta)$ with the approach of \cite{Felder}. 
	%	 Another related question is the study closer the contour deformation non--perturbative definition of the two-cut cubic model \cite{felder} from the viewpoint of both  Painlev\'e II and  the $H_1$ theory

	Finally, concerning the relation between the AD theories and the WKB solutions in terms of holomorphic equations \cite{csm}. Although in the case of cubic and double well potentials we provide the existence of dual AD theories as a justification for the latter, the connection between WKB solutions and holomorphic anomaly equations is expected to be more basic and holds beyond the existence of a dual supersymmetric gauge theory as explained in \cite{csm}. In that perspective it would be interesting to further investigate the algorithm presented in \cite{csm,cms} for more generic quantum mechanical systems with no gauge theory connection.

	\section*{Acknowledgements} 
	
	We would like to thank Giulio Bonelli,  Amir-Kian Kashani-Poor, Vladimir Kazakov,
        Oleg Lisovyy,  Marcos Mari\~{n}o,  Kazunobu Maruyoshi,   Andrei Mironov, Alexei Morozov, Stefano Negro, Antonio
        Sciarappa and Alessandro Tanzini for valuable discussions and
        correspondence.  JG is supported by the European Research
        Council (Programme ``Ideas'' ERC-2012-AdG 320769
        AdS-CFT-solvable).
	
	\appendix
	\raggedbottom
	\section{The $D_k$ coefficients}
	\label{sc:Dk}
	We list here some of the $D_k$ coefficients as computed from the matrix model \footnote{We would like to thank Oleg Lisovyy for sharing with us the unpublished results for $D_4$ and $D_5 $  as computed from the Painlev\'e II equation. We checked that the latters match the ones computed from the matrix model. }
	
	\be \ba  
		D_4(\nu, \theta )&=  \frac{\left(1008845824-45278208 \theta ^2\right) \nu ^{10}}{161243136}+\frac{\left(8989056 \theta ^4-440169984 \theta ^2+4500384000\right) \nu ^8}{161243136} \\
		&+\frac{\left(-793152 \theta ^6+65329920 \theta ^4-1536680448 \theta ^2+8573056768\right) \nu ^6}{161243136}\\
		&+\frac{\left(26244 \theta ^8-3517344 \theta ^6+149049216 \theta ^4-2119879296 \theta ^2+4076024896\right) \nu ^4}{161243136}\\
		&+\frac{\left(32076 \theta ^8-3181680 \theta ^6+97473744 \theta ^4-731835072 \theta ^2+287635968\right) \nu ^2}{161243136}\\
		&+\frac{3267 \theta ^8-338472 \theta ^6+7102512 \theta ^4-29735424 \theta ^2}{161243136}+\frac{83521 \nu ^{12}}{157464}\ .
	\ea\ee
	%%%%
	\be \ba 
		D_5(\nu, \theta)= &-\frac{\ri \left(114050775040-3848647680 \theta ^2\right) \nu ^{13}}{29023764480}\\
		&-\frac{\ri \left(1018759680 \theta ^4-64867691520 \theta ^2+923488394240\right) \nu ^{11}}{29023764480}\\
		&-\frac{\ri \left(-134835840 \theta ^6+13976643840 \theta ^4-437858730240 \theta ^2+3866409244160\right) \nu ^9}{29023764480}\\
		&-\frac{\ri \left(8922960 \theta ^8-1375937280 \theta ^6+72318366720 \theta ^4-1463686214400 \theta ^2+8148525547264\right) \nu ^7}{29023764480}\\
		&-\frac{\ri \left(-236196 \theta ^{10}+56307960 \theta ^8-4676382720 \theta ^6\right) \nu ^5}{29023764480}\\
		&-\frac{\ri \left(167309213760 \theta ^4-2324529313344 \theta ^2+5918924547200\right) \nu ^5}{29023764480}\\
		&-\frac{\ri \left(-481140 \theta ^{10}+85004100 \theta ^8-5528478960 \theta ^6\right) \nu ^3}{29023764480}\\
		&-\frac{\ri \left(150410948640 \theta ^4-1265133321600 \theta ^2+1004626036224\right) \nu ^3}{29023764480}\\
		&-\frac{\ri \left(-147015 \theta ^{10}+26062830 \theta ^8-1548868608 \theta ^6\right) \nu }{29023764480}\\
		&-\frac{\ri \left(30819065760 \theta ^4-151899224832 \theta ^2+12454041600\right) \nu }{29023764480}-\frac{1419857 \ri \nu ^{15}}{7085880}
	\ .\\
	\ea\ee
		
	\be \ba 
		D_6(\nu, \theta)&=\frac{\left(628099301376 \theta ^2-23231463702528\right) \nu ^{16}}{12538266255360}\\
		&+\frac{\left(-207826974720 \theta ^4+16289288110080 \theta ^2-295207026769920\right) \nu ^{14}}{12538266255360}\\
		&+\frac{\left(36675348480 \theta ^6-4596550456320 \theta ^4+179123561349120 \theta ^2-2098472385310720\right) \nu ^{12}}{12538266255360}\\
		&+\frac{\left(-3640567680 \theta ^8+657251573760 \theta ^6-41664329740800 \theta ^4\right) \nu ^{10}}{12538266255360}\\
		&+\frac{\left(1072630221373440 \theta ^2-8861527274778624\right) \nu ^{10}}{12538266255360}\\
		&+\frac{\left(192735936 \theta ^{10}-48583592640 \theta ^8+4552446758400 \theta ^6\right) \nu ^8}{12538266255360}\\
		&+\frac{\left(-195592116614400 \theta ^4+3680491078803456 \theta ^2-20941693135389696\right) \nu ^8}{12538266255360}\\
		&+\frac{\left(-4251528 \theta ^{12}+1617050304 \theta ^{10}-224210592000 \theta ^8+14857834717440 \theta ^6\right) \nu ^6}{12538266255360}\\
		&+\frac{\left(-486010693602432 \theta ^4+6759704932125696 \theta ^2-21152656059375104\right) \nu ^6}{12538266255360}\\
		&+\frac{\left(-12990780 \theta ^{12}+3609891360 \theta ^{10}-399130422480 \theta ^8+21707906542848 \theta ^6\right) \nu ^4}{12538266255360}\\
		&+\frac{\left(-563405245951680 \theta ^4+5235646863930624 \theta ^2-6692702742982656\right) \nu ^4}{12538266255360}\\
		&+\frac{\left(-7938810 \theta ^{12}+2144018160 \theta ^{10}-224279061624 \theta ^8+10830074974272 \theta ^6\right) \nu ^2}{12538266255360}\\
		&+\frac{\left(-215197143901056 \theta ^4+1245163592706048 \theta ^2-361928925941760\right) \nu ^2}{12538266255360}\\
		&+\frac{-539055 \theta ^{12}+157546620 \theta ^{10}-17570163600 \theta ^8+746102664000 \theta ^6}{12538266255360}\\
		&+\frac{-11086283013120 \theta ^4+40217382420480 \theta ^2}{12538266255360}-\frac{24137569 \nu ^{18}}{382637520}
	\ea\ee	
	\be \ba
		D_7(\nu,\theta)&=\frac{\ri \left(2209420712869888-49829211242496 \theta ^2\right) \nu ^{19}}{3159643096350720}+\frac{410338673 \ri \nu ^{21}}{24106163760}\\
		&+\frac{\ri \left(19785127993344 \theta ^4-1841697992687616 \theta ^2+40411929818136576\right) \nu ^{17}}{3159643096350720}\\
		&+\frac{\ri \left(-4364366469120 \theta ^6+642254718099456 \theta ^4-29878724974952448 \theta ^2+431294717560340480\right) \nu ^{15}}{3159643096350720}\\
		&+\frac{\ri \left(-277714792149258240 \theta ^2+2935883132605079552\right) \nu ^{13}}{3159643096350720}\\
		&+\frac{\ri \left(577636738560 \theta ^8-120384194595840 \theta ^6+8952988529479680 \theta ^4\right) \nu ^{13}}{3159643096350720}\\
		&+\frac{\ri \left(+69401564740730880 \theta ^4-1607864190545055744 \theta ^2+12896647995981582336\right) \nu ^{11}}{3159643096350720}\\
		&+\frac{\ri \left(-45871152768 \theta ^{10}+12896334259200 \theta ^8-1378404422115840 \theta ^6\right) \nu ^{11}}{3159643096350720}\\
		&+\frac{\ri \left(+321246107010984960 \theta ^4-5827275251745389568 \theta ^2+34254091748212793344\right) \nu ^9}{3159643096350720}\\
		&+\frac{\ri \left(2023727328 \theta ^{12}-765556795584 \theta ^{10}+113105000211840 \theta ^8-8358622716568320 \theta ^6\right) \nu ^9}{3159643096350720}\\
		&{+}\frac{\ri \left(1707034285647117 \theta ^4-24071708860929693 \theta ^2+88226082084596078\right) \nu ^7}{6171177922560}\\
		&-\frac{\ri  \left(59049 \theta ^8-33332796 \theta ^6+7014457296 \theta ^4-744878628760 \theta ^2+43365988680784\right)\theta ^6 \nu ^7}{4875992432640}\\
		&+\frac{\ri \left(155714834383535 \theta ^4-1576621725969740 \theta ^2+2849012960896848\right) \nu ^5}{391820820480} \\
		& -\frac{\ri  \left(72171 \theta ^8-29238408 \theta ^6+4915342260 \theta ^4-438283228448 \theta ^2+21688585670448\right) \theta ^6\nu ^5}{1393140695040}\\
		&+\frac{\ri \left(107917467861539 \theta ^4-723209858672616 \theta ^2+440213751859680\right) \nu ^3}{457124290560}\\
		&-\frac{\ri  \left(441045 \theta ^8-169143420 \theta ^6+26848410372 \theta ^4-2222362258648 \theta ^2+97346044324736\right)\theta ^6 \nu ^3}{8358844170240}\\
		&+\frac{\ri \left(300249586861 \theta ^4-1289393108560 \theta ^2+81477396000\right) \nu }{8465264640}	\\
		&-\frac{\ri  \left(419265 \theta ^8-168824810 \theta ^6+28035354984 \theta ^4-2301504916992 \theta ^2+88873063642496\right)\theta ^6 \nu }{39007939461120}
	\ea\ee
	 Likewise it is very easy to obtain higher $D_k$, nevertheless the expressions are quite cumbersome and we decided not to write them down explicitly.

\bibliographystyle{JHEP}
\bibliography{biblio}

\providecommand{\href}[2]{#2}\begingroup\raggedright\begin{thebibliography}{10}

\bibitem{ad1}
P.~C. Argyres and M.~R. Douglas, \emph{{New phenomena in SU(3) supersymmetric
  gauge theory}},
  \href{http://dx.doi.org/10.1016/0550-3213(95)00281-V}{\emph{Nucl. Phys.} {\bf
  B448} (1995) 93--126}, [\href{http://arxiv.org/abs/hep-th/9505062}{{\tt
  hep-th/9505062}}].

\bibitem{ad2}
P.~C. Argyres, M.~R. Plesser, N.~Seiberg and E.~Witten, \emph{{New N=2
  superconformal field theories in four-dimensions}},
  \href{http://dx.doi.org/10.1016/0550-3213(95)00671-0}{\emph{Nucl. Phys.} {\bf
  B461} (1996) 71--84}, [\href{http://arxiv.org/abs/hep-th/9511154}{{\tt
  hep-th/9511154}}].

\bibitem{tc12}
T.~Nishinaka and C.~Rim, \emph{{Matrix models for irregular conformal blocks
  and Argyres-Douglas theories}},
  \href{http://dx.doi.org/10.1007/JHEP10(2012)138}{\emph{JHEP} {\bf 10} (2012)
  138}, [\href{http://arxiv.org/abs/1207.4480}{{\tt 1207.4480}}].

\bibitem{Rim:2012tf}
C.~Rim, \emph{{Irregular conformal block and its matrix model}},
  \href{http://arxiv.org/abs/1210.7925}{{\tt 1210.7925}}.

\bibitem{Kajiwara:2004ri}
K.~Kajiwara, T.~Masuda, M.~Noumi, Y.~Ohta and Y.~Yamada, \emph{{Cubic pencils
  and Painleve Hamiltonians}},  \href{http://arxiv.org/abs/nlin/0403009}{{\tt
  nlin/0403009}}.

\bibitem{gil1}
O.~Gamayun, N.~Iorgov and O.~Lisovyy, \emph{{Conformal field theory of
  Painlev\'e VI}}, \href{http://dx.doi.org/10.1007/JHEP10(2012)183,
  10.1007/JHEP10(2012)038}{\emph{JHEP} {\bf 10} (2012) 038},
  [\href{http://arxiv.org/abs/1207.0787}{{\tt 1207.0787}}].

\bibitem{gil}
O.~Gamayun, N.~Iorgov and O.~Lisovyy, \emph{{How instanton combinatorics solves
  Painlev{\'e} VI, V and IIIs}},
  \href{http://dx.doi.org/10.1088/1751-8113/46/33/335203}{\emph{J. Phys.} {\bf
  A46} (2013) 335203}, [\href{http://arxiv.org/abs/1302.1832}{{\tt
  1302.1832}}].

\bibitem{betal}
G.~Bonelli, O.~Lisovyy, K.~Maruyoshi, A.~Sciarappa and A.~Tanzini, \emph{{On
  Painlev\'e/gauge theory correspondence}},
  \href{http://arxiv.org/abs/1612.06235}{{\tt 1612.06235}}.

\bibitem{kkn}
V.~A. Kazakov, I.~K. Kostov and N.~A. Nekrasov, \emph{{D particles, matrix
  integrals and KP hierarchy}},
  \href{http://dx.doi.org/10.1016/S0550-3213(99)00393-4}{\emph{Nucl.Phys.} {\bf
  B557} (1999) 413--442}, [\href{http://arxiv.org/abs/hep-th/9810035}{{\tt
  hep-th/9810035}}].

\bibitem{DiFrancesco:1993cyw}
P.~Di~Francesco, P.~H. Ginsparg and J.~Zinn-Justin, \emph{{2-D Gravity and
  random matrices}},
  \href{http://dx.doi.org/10.1016/0370-1573(94)00084-G}{\emph{Phys. Rept.} {\bf
  254} (1995) 1--133}, [\href{http://arxiv.org/abs/hep-th/9306153}{{\tt
  hep-th/9306153}}].

\bibitem{Cachazo:2001jy}
F.~Cachazo, K.~A. Intriligator and C.~Vafa, \emph{{A Large N duality via a
  geometric transition}},
  \href{http://dx.doi.org/10.1016/S0550-3213(01)00228-0}{\emph{Nucl. Phys.}
  {\bf B603} (2001) 3--41}, [\href{http://arxiv.org/abs/hep-th/0103067}{{\tt
  hep-th/0103067}}].

\bibitem{kmt}
A.~Klemm, M.~Marino and S.~Theisen, \emph{{Gravitational corrections in
  supersymmetric gauge theory and matrix models}},
  \href{http://dx.doi.org/10.1088/1126-6708/2003/03/051}{\emph{JHEP} {\bf 03}
  (2003) 051}, [\href{http://arxiv.org/abs/hep-th/0211216}{{\tt
  hep-th/0211216}}].

\bibitem{kmr}
A.~Klemm, M.~Mari\~no and M.~Rauch, \emph{{Direct Integration and
  Non-Perturbative Effects in Matrix Models}},
  \href{http://dx.doi.org/10.1007/JHEP10(2010)004}{\emph{JHEP} {\bf 10} (2010)
  004}, [\href{http://arxiv.org/abs/1002.3846}{{\tt 1002.3846}}].

\bibitem{huangb}
M.-x. Huang, \emph{{Dijkgraaf-Vafa conjecture and $\beta$-deformed matrix
  models}}, \href{http://dx.doi.org/10.1007/JHEP07(2013)173}{\emph{JHEP} {\bf
  07} (2013) 173}, [\href{http://arxiv.org/abs/1305.1103}{{\tt 1305.1103}}].

\bibitem{Schiappa:2013opa}
R.~Schiappa and R.~Vaz, \emph{{The Resurgence of Instantons: Multi-Cut Stokes
  Phases and the Painleve II Equation}},
  \href{http://dx.doi.org/10.1007/s00220-014-2028-7}{\emph{Commun. Math. Phys.}
  {\bf 330} (2014) 655--721}, [\href{http://arxiv.org/abs/1302.5138}{{\tt
  1302.5138}}].

\bibitem{mmnp}
M.~Marino, \emph{{Nonperturbative effects and nonperturbative definitions in
  matrix models and topological strings}},
  \href{http://dx.doi.org/10.1088/1126-6708/2008/12/114}{\emph{JHEP} {\bf 0812}
  (2008) 114}, [\href{http://arxiv.org/abs/0805.3033}{{\tt 0805.3033}}].

\bibitem{Its:2016jkt}
A.~Its, O.~Lisovyy and A.~Prokhorov, \emph{{Monodromy dependence and connection
  formulae for isomonodromic tau functions}},
  \href{http://arxiv.org/abs/1604.03082}{{\tt 1604.03082}}.

\bibitem{csm}
S.~Codesido and M.~Marino, \emph{{Holomorphic Anomaly and Quantum Mechanics}},
  \href{http://dx.doi.org/10.1088/1751-8121/aa9e77}{\emph{J. Phys.} {\bf A51}
  (2018) 055402}, [\href{http://arxiv.org/abs/1612.07687}{{\tt 1612.07687}}].

\bibitem{Dijkgraaf:2002pp}
R.~Dijkgraaf, S.~Gukov, V.~A. Kazakov and C.~Vafa, \emph{{Perturbative analysis
  of gauged matrix models}},
  \href{http://dx.doi.org/10.1103/PhysRevD.68.045007}{\emph{Phys. Rev.} {\bf
  D68} (2003) 045007}, [\href{http://arxiv.org/abs/hep-th/0210238}{{\tt
  hep-th/0210238}}].

\bibitem{Maruyoshi:2014eja}
K.~Maruyoshi, \emph{{$\beta$-deformed matrix models and the 2d/4d
  correspondence}},  in \emph{New Dualities of Supersymmetric Gauge Theories}
  (J.~Teschner, ed.), pp.~121--157.
\newblock 2016.
\newblock \href{http://arxiv.org/abs/1412.7124}{{\tt 1412.7124}}.
\newblock \href{http://dx.doi.org/10.1007/978-3-319-18769-3_5}{DOI}.

\bibitem{bmt}
G.~Bonelli, K.~Maruyoshi and A.~Tanzini, \emph{{Quantum Hitchin Systems via
  beta-deformed Matrix Models}},  \href{http://arxiv.org/abs/1104.4016}{{\tt
  1104.4016}}.

\bibitem{Masuda:1996xj}
T.~Masuda and H.~Suzuki, \emph{{Periods and prepotential of N=2 SU(2)
  supersymmetric Yang-Mills theory with massive hypermultiplets}},
  \href{http://dx.doi.org/10.1142/S0217751X97001791}{\emph{Int. J. Mod. Phys.}
  {\bf A12} (1997) 3413--3431},
  [\href{http://arxiv.org/abs/hep-th/9609066}{{\tt hep-th/9609066}}].

\bibitem{Bonelli:2011aa}
G.~Bonelli, K.~Maruyoshi and A.~Tanzini, \emph{{Wild Quiver Gauge Theories}},
  \href{http://dx.doi.org/10.1007/JHEP02(2012)031}{\emph{JHEP} {\bf 02} (2012)
  031}, [\href{http://arxiv.org/abs/1112.1691}{{\tt 1112.1691}}].

\bibitem{Gaiotto:2012sf}
D.~Gaiotto and J.~Teschner, \emph{{Irregular singularities in Liouville theory
  and Argyres-Douglas type gauge theories, I}},
  \href{http://dx.doi.org/10.1007/JHEP12(2012)050}{\emph{JHEP} {\bf 12} (2012)
  050}, [\href{http://arxiv.org/abs/1203.1052}{{\tt 1203.1052}}].

\bibitem{Akemann:1996zr}
G.~Akemann, \emph{{Higher genus correlators for the Hermitian matrix model with
  multiple cuts}},
  \href{http://dx.doi.org/10.1016/S0550-3213(96)00542-1}{\emph{Nucl. Phys.}
  {\bf B482} (1996) 403--430}, [\href{http://arxiv.org/abs/hep-th/9606004}{{\tt
  hep-th/9606004}}].

\bibitem{bcov}
M.~Bershadsky, S.~Cecotti, H.~Ooguri and C.~Vafa, \emph{{Holomorphic anomalies
  in topological field theories}},
  \href{http://dx.doi.org/10.1016/0550-3213(93)90548-4}{\emph{Nucl.Phys.} {\bf
  B405} (1993) 279--304}, [\href{http://arxiv.org/abs/hep-th/9302103}{{\tt
  hep-th/9302103}}].

\bibitem{eo}
B.~Eynard and N.~Orantin, \emph{{Invariants of algebraic curves and topological
  expansion}},
  \href{http://dx.doi.org/10.4310/CNTP.2007.v1.n2.a4}{\emph{Commun.Num.Theor.Phys.}
  {\bf 1} (2007) 347--452}, [\href{http://arxiv.org/abs/math-ph/0702045}{{\tt
  math-ph/0702045}}].

\bibitem{hk}
M.-x. Huang and A.~Klemm, \emph{{Direct integration for general $\Omega$
  backgrounds}},
  \href{http://dx.doi.org/10.4310/ATMP.2012.v16.n3.a2}{\emph{Adv.Theor.Math.Phys.}
  {\bf 16} (2012) 805--849}, [\href{http://arxiv.org/abs/1009.1126}{{\tt
  1009.1126}}].

\bibitem{kwal}
D.~Krefl and J.~Walcher, \emph{{Extended Holomorphic Anomaly in Gauge Theory}},
  \href{http://dx.doi.org/10.1007/s11005-010-0432-2}{\emph{Lett. Math. Phys.}
  {\bf 95} (2011) 67--88}, [\href{http://arxiv.org/abs/1007.0263}{{\tt
  1007.0263}}].

\bibitem{n}
N.~A. Nekrasov, \emph{{Seiberg-Witten prepotential from instanton counting}},
  \href{http://dx.doi.org/10.4310/ATMP.2003.v7.n5.a4}{\emph{Adv.Theor.Math.Phys.}
  {\bf 7} (2004) 831--864}, [\href{http://arxiv.org/abs/hep-th/0206161}{{\tt
  hep-th/0206161}}].

\bibitem{Nekrasov:2003rj}
N.~Nekrasov and A.~Okounkov, \emph{{Seiberg-Witten theory and random
  partitions}}, \href{http://dx.doi.org/10.1007/0-8176-4467-9_15}{\emph{Prog.
  Math.} {\bf 244} (2006) 525--596},
  [\href{http://arxiv.org/abs/hep-th/0306238}{{\tt hep-th/0306238}}].

\bibitem{Huang:2006si}
M.-x. Huang and A.~Klemm, \emph{{Holomorphic Anomaly in Gauge Theories and
  Matrix Models}},
  \href{http://dx.doi.org/10.1088/1126-6708/2007/09/054}{\emph{JHEP} {\bf 09}
  (2007) 054}, [\href{http://arxiv.org/abs/hep-th/0605195}{{\tt
  hep-th/0605195}}].

\bibitem{Huang:2009md}
M.-x. Huang and A.~Klemm, \emph{{Holomorphicity and Modularity in
  Seiberg-Witten Theories with Matter}},
  \href{http://dx.doi.org/10.1007/JHEP07(2010)083}{\emph{JHEP} {\bf 07} (2010)
  083}, [\href{http://arxiv.org/abs/0902.1325}{{\tt 0902.1325}}].

\bibitem{bgt}
G.~Bonelli, A.~Grassi and A.~Tanzini, \emph{{Seiberg--Witten theory as a Fermi
  gas}}, \href{http://dx.doi.org/10.1007/s11005-017-0945-z,
  10.1007/s11005-016-0893-z}{\emph{Lett. Math. Phys.} {\bf 107} (2017) 1--30},
  [\href{http://arxiv.org/abs/1603.01174}{{\tt 1603.01174}}].

\bibitem{bgt2}
G.~Bonelli, A.~Grassi and A.~Tanzini, \emph{{New results in $\mathcal{N}=2$
  theories from non-perturbative string}},
  \href{http://arxiv.org/abs/1704.01517}{{\tt 1704.01517}}.

\bibitem{mz}
M.~Marino and S.~Zakany, \emph{{Matrix models from operators and topological
  strings}}, \href{http://dx.doi.org/10.1007/s00023-015-0422-0}{\emph{Annales
  Henri Poincare} {\bf 17} (2016) 1075--1108},
  [\href{http://arxiv.org/abs/1502.02958}{{\tt 1502.02958}}].

\bibitem{kmz}
R.~Kashaev, M.~Marino and S.~Zakany, \emph{{Matrix models from operators and
  topological strings, 2}},
  \href{http://dx.doi.org/10.1007/s00023-016-0471-z}{\emph{Annales Henri
  Poincare} {\bf 17} (2016) 2741--2781},
  [\href{http://arxiv.org/abs/1505.02243}{{\tt 1505.02243}}].

\bibitem{cgum}
S.~Codesido, J.~Gu and M.~Marino, \emph{{Operators and higher genus mirror
  curves}}, \href{http://dx.doi.org/10.1007/JHEP02(2017)092}{\emph{JHEP} {\bf
  02} (2017) 092}, [\href{http://arxiv.org/abs/1609.00708}{{\tt 1609.00708}}].

\bibitem{cgm2}
S.~Codesido, A.~Grassi and M.~Marino, \emph{{Spectral Theory and Mirror Curves
  of Higher Genus}},
  \href{http://dx.doi.org/10.1007/s00023-016-0525-2}{\emph{Annales Henri
  Poincare} {\bf 18} (2017) 559--622},
  [\href{http://arxiv.org/abs/1507.02096}{{\tt 1507.02096}}].

\bibitem{Bonelli:2017gdk}
G.~Bonelli, A.~Grassi and A.~Tanzini, \emph{{Quantum curves and $q$-deformed
  Painlev\'e equations}},  \href{http://arxiv.org/abs/1710.11603}{{\tt
  1710.11603}}.

\bibitem{Mironov:2017lgl}
A.~Mironov and A.~Morozov, \emph{{On determinant representation and
  integrability of Nekrasov functions}},
  \href{http://dx.doi.org/10.1016/j.physletb.2017.08.004}{\emph{Phys. Lett.}
  {\bf B773} (2017) 34--46}, [\href{http://arxiv.org/abs/1707.02443}{{\tt
  1707.02443}}].

\bibitem{Mironov:2017sqp}
A.~Mironov and A.~Morozov, \emph{{q-Painleve equation from Virasoro
  constraints}},  \href{http://arxiv.org/abs/1708.07479}{{\tt 1708.07479}}.

\bibitem{ilt}
A.~Its, O.~Lisovyy and {\relax Yu}.~Tykhyy, \emph{{Connection problem for the
  sine-Gordon/Painlev\'e III tau function and irregular conformal blocks}},
  {\emph{Int. Math. Res. Notices} {\bf 18} (2015) 8903--8924},
  [\href{http://arxiv.org/abs/1403.1235}{{\tt 1403.1235}}].

\bibitem{msw}
M.~Marino, R.~Schiappa and M.~Weiss, \emph{{Multi-Instantons and Multi-Cuts}},
  \href{http://dx.doi.org/10.1063/1.3097755}{\emph{J.Math.Phys.} {\bf 50}
  (2009) 052301}, [\href{http://arxiv.org/abs/0809.2619}{{\tt 0809.2619}}].

\bibitem{bde}
G.~Bonnet, F.~David and B.~Eynard, \emph{{Breakdown of universality in multicut
  matrix models}},
  \href{http://dx.doi.org/10.1088/0305-4470/33/38/307}{\emph{J.Phys.} {\bf A33}
  (2000) 6739--6768}, [\href{http://arxiv.org/abs/cond-mat/0003324}{{\tt
  cond-mat/0003324}}].

\bibitem{em}
B.~Eynard and M.~Marino, \emph{{A Holomorphic and background independent
  partition function for matrix models and topological strings}},
  \href{http://dx.doi.org/10.1016/j.geomphys.2010.11.012}{\emph{J.Geom.Phys.}
  {\bf 61} (2011) 1181--1202}, [\href{http://arxiv.org/abs/0810.4273}{{\tt
  0810.4273}}].

\bibitem{eynard}
B.~Eynard, \emph{{Large N expansion of convergent matrix integrals, holomorphic
  anomalies, and background independence}},
  \href{http://dx.doi.org/10.1088/1126-6708/2009/03/003}{\emph{JHEP} {\bf 0903}
  (2009) 003}, [\href{http://arxiv.org/abs/0802.1788}{{\tt 0802.1788}}].

\bibitem{Felder}
G.~Felder and R.~Riser, \emph{{Holomorphic matrix integrals}},
  \href{http://dx.doi.org/10.1016/j.nuclphysb.2004.05.010}{\emph{Nucl. Phys.}
  {\bf B691} (2004) 251--258}, [\href{http://arxiv.org/abs/hep-th/0401191}{{\tt
  hep-th/0401191}}].

\bibitem{ns}
N.~A. Nekrasov and S.~L. Shatashvili, \emph{{Quantization of Integrable Systems
  and Four Dimensional Gauge Theories}},
  \href{http://arxiv.org/abs/0908.4052}{{\tt 0908.4052}}.

\bibitem{ggu}
A.~Grassi and J.~Gu, \emph{{BPS relations from spectral problems and blowup
  equations}},  \href{http://arxiv.org/abs/1609.05914}{{\tt 1609.05914}}.

\bibitem{mirmor2}
A.~Mironov and A.~Morozov, \emph{{Nekrasov Functions from Exact BS Periods: The
  Case of SU(N)}},
  \href{http://dx.doi.org/10.1088/1751-8113/43/19/195401}{\emph{J.Phys.} {\bf
  A43} (2010) 195401}, [\href{http://arxiv.org/abs/0911.2396}{{\tt
  0911.2396}}].

\bibitem{Basar:2017hpr}
G.~Basar, G.~V. Dunne and M.~Unsal, \emph{{Quantum Geometry of Resurgent
  Perturbative/Nonperturbative Relations}},
  \href{http://dx.doi.org/10.1007/JHEP05(2017)087}{\emph{JHEP} {\bf 05} (2017)
  087}, [\href{http://arxiv.org/abs/1701.06572}{{\tt 1701.06572}}].

\bibitem{mirmor}
A.~Mironov and A.~Morozov, \emph{{Nekrasov Functions and Exact Bohr-Zommerfeld
  Integrals}}, \href{http://dx.doi.org/10.1007/JHEP04(2010)040}{\emph{JHEP}
  {\bf 1004} (2010) 040}, [\href{http://arxiv.org/abs/0910.5670}{{\tt
  0910.5670}}].

\bibitem{acdkv}
M.~Aganagic, M.~C. Cheng, R.~Dijkgraaf, D.~Krefl and C.~Vafa, \emph{{Quantum
  Geometry of Refined Topological Strings}},
  \href{http://dx.doi.org/10.1007/JHEP11(2012)019}{\emph{JHEP} {\bf 1211}
  (2012) 019}, [\href{http://arxiv.org/abs/1105.0630}{{\tt 1105.0630}}].

\bibitem{Nekrasov:1996cz}
N.~Nekrasov, \emph{{Five dimensional gauge theories and relativistic integrable
  systems}}, \href{http://dx.doi.org/10.1016/S0550-3213(98)00436-2}{\emph{Nucl.
  Phys.} {\bf B531} (1998) 323--344},
  [\href{http://arxiv.org/abs/hep-th/9609219}{{\tt hep-th/9609219}}].

\bibitem{ghm}
A.~Grassi, Y.~Hatsuda and M.~Marino, \emph{{Topological Strings from Quantum
  Mechanics}}, \href{http://dx.doi.org/10.1007/s00023-016-0479-4}{\emph{Annales
  Henri Poincare} {\bf 17} (2016) 3177--3235},
  [\href{http://arxiv.org/abs/1410.3382}{{\tt 1410.3382}}].

\bibitem{gkmr}
J.~Gu, A.~Klemm, M.~Marino and J.~Reuter, \emph{{Exact solutions to quantum
  spectral curves by topological string theory}},
  \href{http://dx.doi.org/10.1007/JHEP10(2015)025}{\emph{JHEP} {\bf 10} (2015)
  025}, [\href{http://arxiv.org/abs/1506.09176}{{\tt 1506.09176}}].

\bibitem{Wang:2015wdy}
X.~Wang, G.~Zhang and M.-x. Huang, \emph{{New Exact Quantization Condition for
  Toric Calabi-Yau Geometries}},
  \href{http://dx.doi.org/10.1103/PhysRevLett.115.121601}{\emph{Phys. Rev.
  Lett.} {\bf 115} (2015) 121601}, [\href{http://arxiv.org/abs/1505.05360}{{\tt
  1505.05360}}].

\bibitem{huang1606}
K.~Sun, X.~Wang and M.-x. Huang, \emph{{Exact Quantization Conditions, Toric
  Calabi-Yau and Nonperturbative Topological String}},
  \href{http://dx.doi.org/10.1007/JHEP01(2017)061}{\emph{JHEP} {\bf 01} (2017)
  061}, [\href{http://arxiv.org/abs/1606.07330}{{\tt 1606.07330}}].

\bibitem{Grassi:2017qee}
A.~Grassi and M.~Marino, \emph{{The complex side of the TS/ST correspondence}},
   \href{http://arxiv.org/abs/1708.08642}{{\tt 1708.08642}}.

\bibitem{Dunham1932:WKB}
J.~Dunham, \emph{{The Wentzel-Brillouin-Kramers method of solving the wave
  equation}}, {\emph{Phys. Rev.} {\bf 41} (1932) 713--720}.

\bibitem{Bender1977:WKB}
C.~M. Bender, K.~Olaussen and P.~Wang, \emph{{Numerological analysis of the WKB
  approximation in large order}}, {\emph{Phys. Rev. D} {\bf 16} (1977)
  1740--1748}.

\bibitem{Galindo1990:WKB}
A.~Galindo and P.~Pascual, \emph{{Quantum Mechanics}}, vol.~2.
\newblock Springer-Verlag, 1990.

\bibitem{voross}
A.~Voros, \emph{{The return of the quartic oscillator. The complex WKB
  method}}, {\emph{Ann. Inst. H. Poincar\'e} {\bf A 39} (1983) 211}.

\bibitem{Voros1981}
A.~Voros, \emph{{Spectre de l'\'{e}quation de Schr\"{o}dinger et m\'{e}thode
  BKW}}.
\newblock Publications Math\'{e}matiques d'Orsay, 1981.

\bibitem{Silverstone1985:WKB}
H.~J. Silverstone, \emph{{JWKB connection-formula problem revisited via Borel
  summation}}, {\emph{Phys. Rev. Lett.} {\bf 55} (1985) 2523}.

\bibitem{cms}
S.~Codesido, M.~Marino and R.~Schiappa, \emph{{Non-Perturbative Quantum
  Mechanics from Non-Perturbative Strings}},
  \href{http://arxiv.org/abs/1712.02603}{{\tt 1712.02603}}.

\bibitem{cesv1}
R.~Couso-Santamar{\'\i}a, J.~D. Edelstein, R.~Schiappa and M.~Vonk,
  \emph{{Resurgent Transseries and the Holomorphic Anomaly}},
  \href{http://dx.doi.org/10.1007/s00023-015-0407-z}{\emph{Annales Henri
  Poincare} {\bf 17} (2016) 331--399},
  [\href{http://arxiv.org/abs/1308.1695}{{\tt 1308.1695}}].

\bibitem{ag}
A.~Grassi, \emph{{Spectral determinants and quantum theta functions}},
  \href{http://dx.doi.org/10.1088/1751-8113/49/50/505401}{\emph{J. Phys.} {\bf
  A49} (2016) 505401}, [\href{http://arxiv.org/abs/1604.06786}{{\tt
  1604.06786}}].

\bibitem{ito-shu}
K.~Ito and H.~Shu, \emph{{ODE/IM correspondence and the Argyres-Douglas
  theory}}, \href{http://dx.doi.org/10.1007/JHEP08(2017)071}{\emph{JHEP} {\bf
  08} (2017) 071}, [\href{http://arxiv.org/abs/1707.03596}{{\tt 1707.03596}}].

\bibitem{Dorey:2007zx}
P.~Dorey, C.~Dunning and R.~Tateo, \emph{{The ODE/IM Correspondence}},
  \href{http://dx.doi.org/10.1088/1751-8113/40/32/R01}{\emph{J. Phys.} {\bf
  A40} (2007) R205}, [\href{http://arxiv.org/abs/hep-th/0703066}{{\tt
  hep-th/0703066}}].

\bibitem{Dorey:1998pt}
P.~Dorey and R.~Tateo, \emph{{Anharmonic oscillators, the thermodynamic Bethe
  ansatz, and nonlinear integral equations}},
  \href{http://dx.doi.org/10.1088/0305-4470/32/38/102}{\emph{J. Phys.} {\bf
  A32} (1999) L419--L425}, [\href{http://arxiv.org/abs/hep-th/9812211}{{\tt
  hep-th/9812211}}].

\bibitem{Douglas:1989ve}
M.~R. Douglas and S.~H. Shenker, \emph{{Strings in Less Than One-Dimension}},
  \href{http://dx.doi.org/10.1016/0550-3213(90)90522-F}{\emph{Nucl. Phys.} {\bf
  B335} (1990) 635}.

\bibitem{Brezin:1990rb}
E.~Brezin and V.~A. Kazakov, \emph{{Exactly Solvable Field Theories of Closed
  Strings}}, \href{http://dx.doi.org/10.1016/0370-2693(90)90818-Q}{\emph{Phys.
  Lett.} {\bf B236} (1990) 144--150}.

\bibitem{Gross:1989vs}
D.~J. Gross and A.~A. Migdal, \emph{{Nonperturbative Two-Dimensional Quantum
  Gravity}}, \href{http://dx.doi.org/10.1103/PhysRevLett.64.127}{\emph{Phys.
  Rev. Lett.} {\bf 64} (1990) 127}.

\bibitem{Kharchev:1991cy}
S.~Kharchev, A.~Marshakov, A.~Mironov, A.~Morozov and A.~Zabrodin,
  \emph{{Towards unified theory of 2-d gravity}},
  \href{http://dx.doi.org/10.1016/0550-3213(92)90521-C}{\emph{Nucl. Phys.} {\bf
  B380} (1992) 181--240}, [\href{http://arxiv.org/abs/hep-th/9201013}{{\tt
  hep-th/9201013}}].

\bibitem{dss}
M.~R. Douglas, N.~Seiberg and S.~H. Shenker, \emph{Flow and instability in
  quantum gravity}, {\emph{Phys. Lett. B} {\bf 244} (1990) 381--386}.

\bibitem{cmo}
C.Crnkovi\'c and G.~Moore, \emph{Multicritical multi-cut matrix models},
  {\emph{Phys. Lett. B} {\bf 257} (1991) }.

\bibitem{Mironov:1994mv}
A.~Mironov, A.~Morozov and G.~W. Semenoff, \emph{{Unitary matrix integrals in
  the framework of generalized Kontsevich model. 1. Brezin-Gross-Witten
  model}}, \href{http://dx.doi.org/10.1142/S0217751X96002339}{\emph{Int. J.
  Mod. Phys.} {\bf A11} (1996) 5031--5080},
  [\href{http://arxiv.org/abs/hep-th/9404005}{{\tt hep-th/9404005}}].

\bibitem{Lisovyy:2016qig}
O.~Lisovyy and J.~Roussillon, \emph{{On the connection problem for Painlev{\'e}
  I}}, \href{http://dx.doi.org/10.1088/1751-8121/aa6e12}{\emph{J. Phys.} {\bf
  A50} (2017) 255202}, [\href{http://arxiv.org/abs/1612.08382}{{\tt
  1612.08382}}].

\bibitem{Cordova:2016jlu}
C.~Cordova, B.~Heidenreich, A.~Popolitov and S.~Shakirov, \emph{{Orbifolds and
  Exact Solutions of Strongly-Coupled Matrix Models}},
  \href{http://arxiv.org/abs/1611.03142}{{\tt 1611.03142}}.

\end{thebibliography}\endgroup
\end{document}